\newif\ifsubmode
\submodefalse

\newif\ifprintfig
\printfigtrue

\ifsubmode
  \documentclass[12pt,preprint]{aastex}
  \received{} 
  \revised{}
  \accepted{}
  \ccc{}
  \cpright{}{}
\else
  \documentclass[preprint]{aastex}
  \usepackage{epsf}
  \slugcomment{{\it draft \today, do not distribute}}
\fi

\shortauthors{Verdoes Kleijn et al.}
\shorttitle{The Nuclear Gas Dispersion in Early-Type Galaxies}

\newcommand{\etal}{{et al.\ }}

\newcommand{\lta}{\lesssim}
\newcommand{\gta}{\gtrsim}
\renewcommand{\deg}{^{\circ}}

\newcommand{\kms}{\>{\rm km}\,{\rm s}^{-1}}

\newcommand{\pc}{\>{\rm pc}}
\newcommand{\Mpc}{\>{\rm Mpc}}
\newcommand{\Msun}{\>{\rm M_{\odot}}}

\newcommand{\Mbh}{M_{\bullet}}

\newcommand{\HalphaNII}{H$\alpha$+[NII]}

\begin{document}

\title{Understanding the Nuclear Gas Dispersion in Early-Type Galaxies in the
Context of Black Hole Demographics\altaffilmark{1}}

\author{%
Gijs A.\ Verdoes Kleijn,\altaffilmark{2,3}
Roeland P.\ van der Marel,\altaffilmark{4}
Jacob Noel-Storr\altaffilmark{4,5} \\
}


\altaffiltext{1}{Based on observations with the NASA/ESA Hubble Space 
       Telescope obtained at the Space Telescope Science Institute, which is 
       operated by the Association of Universities for Research in Astronomy, 
       Incorporated, under NASA contract NAS5-26555.}

\altaffiltext{2}{ESO, Karl-Schwarzschild-Strasse 2, 85748, Garching bei Muenchen, Germany}

\altaffiltext{3}{current address: Kapteyn Astronomical Institute, Groningen, 9700 AV, The Netherlands }

\altaffiltext{4}{Space Telescope Science Institute, 3700 San Martin Drive, 
Baltimore, MD 21218.}

\altaffiltext{5}{current address: Steward Observatory, University of Arizona, 933 N Cherry Ave., Tucson, AZ 85721}


\ifsubmode\else
\clearpage\fi


\ifsubmode\else
\baselineskip=14pt
\fi


\begin{abstract}
The majority of nearby early-type galaxies contains detectable amounts
of emission-line gas at their centers. The nuclear gas kinematics form
a valuable diagnostic of the central black hole (BH) mass. Here we
analyze and model HST/STIS observations of a sample of 27 galaxies; 16
Fanaroff \& Riley Type I radio galaxies and 11 (more) normal
early-type galaxies. We focus here on what can be learned from
the nuclear velocity dispersion (line width) of the gas as a complement to the many studies dealing with gas rotation velocities. We find that
the dispersion in a STIS aperture of $\sim 0.1''$--$0.2''$ generally
exceeds the large-scale stellar velocity dispersion of the galaxy.
This is qualitatively consistent with the presence of central BHs, but
raises the questions whether the excess gas dispersion is of
gravitational or non-gravitational origin and whether the implied BH
masses are consistent with our current understanding of BH demography
(as predicted by the $M-\sigma$ relation between BH mass and stellar
velocity dispersion). To address this we construct purely
gravitational axisymmetric dynamical models for the gas, both thin
disk models and models with more general axis ratios and velocity
anisotropies. For the normal galaxies the nuclear gas dispersions are
adequately reproduced assuming disks around BHs with masses that follow
the $M-\sigma$ relation. In contrast, the gas dispersions observed for the
radio galaxies generally exceed those predicted by any of the models. We
attribute this to the presence of non-gravitational motions in the gas
that are similar to or larger than the gravitational motions. The
non-gravitational motions are presumably driven by the active galactic
nucleus (AGN), but we do not find a relation between the radiative
output of the AGN and the non-gravitational dispersion. Given the
uncertainties about the dynamical state of the gas, it is not possible
to uniquely determine the BH mass for each galaxy from its nuclear gas
dispersion. However, for the sample as a whole the observed
dispersions do not provide evidence for significant deviations from
the $M-\sigma$ relation.
\end{abstract}



\clearpage


\section{Introduction}
\label{s:intro}

The mass of a central black hole (BH) in a galaxy can be directly
weighed using gravitational 'test particles' moving around it (see
e.g., Kormendy \& Richstone 1995). Commonly used test particles
are stars, optical emission-line gas and maser clouds. These move
around in the combined potential well of the stellar mass and the
central black hole mass.  A secure direct dynamical measurement of the
BH mass requires that the distribution and kinematics of the test
particles are measured in the immediate vicinity of the BH, i.e., the
'black hole's sphere of influence', so that its gravitational
potential has a measurable effect on the kinematics in addition to the
effect of the stellar mass potential. Ground-based optical telescopes
can resolve the BH's sphere of influence only for very nearby
galaxies, with our Milky Way being a spectacular nearby example (e.g.,
Ghez \etal 2003; Sch{\"o}del \etal 2003). The Hubble Space Telescope
(HST) can provide the required spatial resolution for galaxies with
distances up to several tens of Mpc (e.g., Kormendy \& Gebhardt
2001). The Very Long Baseline Interferometer (VLBI) can in
principle probe BH masses out to even larger distances provided
that the galaxies display nuclear maser emission. Together these
methods have led to direct dynamical measurements of BH masses $\Mbh$
in the range $\Mbh \sim 10^6 - 5 \times 10^9 \Msun$ in several tens of
galaxies in the nearby Universe (see e.g., Tremaine \etal 2002;
Marconi \etal 2003 for listings of BH mass measurements). The
BH masses are mostly based on dynamical modeling of kinematics of
either stars or emission-line gas. The detected black hole masses
$\Mbh$ correlate well with both host spheroid luminosity $L_B$, global
stellar velocity dispersion $\sigma_{\rm s}$ and spheroid mass
(Gebhardt et al., 2000; Ferrarese \& Merritt, 2000; Tremaine
\etal 2002; Marconi \etal 2003; H\"aring \& Rix 2004). 
These correlations roughly represent our current knowledge of the
local BH demography.  To explore these correlations further, i.e.,
their scatter and extent in BH mass, dynamical modeling using
emission-line gas kinematics in early-type galaxies is convenient and
in some respects even crucial. First of all, more than 50\% of the
early-type galaxies in the nearby Universe contain gas at their
centers (e.g., Goudfrooij \etal 1994; Ho, Filippenko \& Sargent
1997). Second of all, the most massive black holes currently found
(i.e., $\Mbh \gta 10^9 \Msun$) are often in giant ellipticals. The
central stellar surface brightness of these galaxies is
typically too low to obtain accurate stellar kinematics with HST
and maser emission has not been detected (e.g., Barth
\etal 2004).

At the centers of galaxies, the collisional gas is expected to settle
quickly into a disk, if unperturbed by forces other than
gravity (e.g., Habe \& Ikeuchi 1985). The frequent detection
with HST of disk-like structures in dust and gas in
early-type galaxies suggests that settling actually takes place (e.g.,
van Dokkum \& Franx 1995; Tran \etal 2001; Verdoes Kleijn \etal
1999; Laine \etal 2003). Therefore the extended gas
velocities in galactic nuclei have been modeled assuming a thin disk
in circular rotation. Such models have been successful in explaining
the observed gas velocities.  However, in the few cases where both gas
velocities and gas dispersions are modeled, it is often found that the
nuclear velocity dispersion exceeds the prediction for a thin disk
(van der Marel \& van den Bosch
1998; Barth \etal 2001; Maciejewski \& Binney 2001; Cappellari et al.~2002; Verdoes Kleijn et al.~2002). Many of
these targets are AGN and display regular disks of dust and
gas.  The origin of this 'excess' velocity dispersion is
unknown. Two common ad-hoc assumptions are that: (a) the
excess dispersion is due to non-gravitational 'turbulent' forces of
unknown origin which do not affect the mean circular rotation of the
gas (e.g., van der Marel \& van den Bosch 1998; Verdoes Kleijn \etal 2002); or (b) the excess dispersion is purely gravitational
and affects the rotation of the gas through the asymmetric drift
equation (e.g., Barth \etal 2001).

A main aim of this paper is to determine which early-type galaxies do and which do not display an excess in
nuclear gas dispersion.  If there are galaxies for which the disk remains thin all the way to the nucleus,
then any velocity dispersion observed through a nuclear aperture is caused by differential rotation over this
aperture. A nuclear aperture samples gas which is much closer to the BH than a series of apertures which
samples the extended rotation curve. Hence, the nuclear gas dispersion can be sensitive to much lower black
hole masses than the rotation curve. This is relevant as the sensitivity in terms of the minimum detectable
BH mass in early-type galaxies with stellar dynamical methods or with gas rotation velocities is typically
not much below the values predicted by the correlations between BH mass and spheroid properties. A relevant
example in this case is the claim that active spiral galaxies classified as Narrow Line Seyfert 1s harbor BHs
with masses smaller than those predicted by the $\Mbh-\sigma$ relation (e.g., Grupe \& Mathur 2004).

For the galaxies with nuclear gas dispersions in excess of that
expected from a thin disk, we want to determine the origin of this
excess.  On the one hand, if it has a gravitational origin, this
implies that the gas should rotate less fast than the circular speed
and the thin-disk approximation does not hold. The gas might
then have a distribution somewhere in the range from a thin
disk to a purely spherical distribution. Previous BH mass
estimates based on the assumption of thin disks would then
be underestimates, and this would affect the observed
correlations between black hole mass and large-scale spheroid
properties.  On the other hand, if the excess dispersion has a
non-gravitational origin, we would like to quantify how it depends on galaxy parameters, such as nuclear
activity. This in turn could shed light on the accretion process in
these nearby, typically low-luminosity, AGN.

In summary, this paper studies the nature of the line widths of
emission-gas in nearby early-type galaxy nuclei and its implications for
the BH demography and BH accretion processes. A similar study into the
line-widths of gas for late-type galaxies has been performed by Sarzi
\etal (2002).  The outline of the paper is as
follows. Section~\ref{s:sampledata} describes the sample selection,
the data and the determination of the central gas velocity
dispersions. Section~\ref{s:excess} compares the observed dispersions
to those expected from a thin circular disk. Section~\ref{s:spheroid}
compares the observed dispersions to those expected from a gas
distribution which is more spheroidal than a
disk. Section~\ref{s:nongrav} constrains the direction of the
non-gravitational dispersion. Section~\ref{s:cave} describes
the caveats of our modeling. Finally, Section~\ref{s:conc}
provides a discussion of the results and lists the main conclusions of
the paper. Throughout the paper we use a Hubble constant $H_0$=
75 kms$^{-1}$Mpc$^{-1}$.

\section{Sample and Data Analysis}
\label{s:sampledata}

We analyze a sample of 27 galaxies. It consists of
galaxies which meet the following requirements:  (a) early-type
host morphology; (b) distance below 110 Mpc; (c) a large-scale stellar
velocity dispersion measurement is available from the literature; (d)
HST imaging exists that shows an identifiable (i.e., relatively
unobscured) nucleus; (e) HST emission line spectroscopy exists for
which kinematics and flux profiles were published (20 galaxies) or
which are available in the HST archive (7 galaxies). We excluded
galaxies for which the signal-to-noise ratio of the spectrum
was very low, complicating a reliable analysis. Sixteen of the
galaxies are classified as Fanaroff \& Riley Type I radio galaxies
with jets on scales of tens of kpc or larger. The other 11 galaxies
also have radio emission. Their radio luminosity is typically much
lower (cf.~Table~\ref{t:sample}) and the radio emission originates
from a central compact component. We reduced and
analyzed the archival data for the 7 galaxies similarly as
in our previous papers (e.g., Noel-Storr
\etal 2003; Verdoes Kleijn \etal 2002). The spectra for all galaxies
except two were obtained with HST/STIS in combination with the G750M
grating. For IC 1459 a HST/STIS spectrum with the G430L grating was
obtained and for NGC 6251 a HST/FOS spectrum with the $0.1''$ PAIR B
aperture. All spectra cover the H$\alpha$+[NII] and
[SII]6716,6731 lines, except for the G430L spectrum of IC
1459. In that case we use the H$\beta$ line for our
analysis. Table~\ref{t:sample} lists basic data for the sample
galaxies.

The first step is to determine the gas velocity dispersions in the
central region of each galaxy. In most cases (both in the
literature and in our analysis of the archival spectra) single
Gaussian fits were made to each of the lines of the
{\HalphaNII} composite and to the weaker [SII]6716, 6731 lines. Only
in NGC 3245 and NGC 4526 are the [SII] lines too weak to be
fitted. For NGC 6251 we used the fit to the {\HalphaNII}
lines available in Ferrarese \& Ford (1999). For these narrow components, the [NII]$\lambda6584$ was the strongest line in $\sim 85\%$ of the galaxies. In four galaxies (NGC
3245, NGC 3998, NGC 4278 and NGC 6251) a broad emission-line component
is clearly present in addition to the narrow {\HalphaNII} components. This is most likely due to a broad H$\alpha$
component because the broad component is not seen in the [SII]
forbidden lines and the forbidden [NII] lines originate presumably in
the same region as the [SII] lines. This broad component
was fitted with an additional Gaussian. For the two archival
sources with a broad component (NGC 3998 and NGC 4278) we
established that the [SII] lines provide an important fitting
constraint in addition to the {\HalphaNII} region which is blended
heavily due to the broad component. Excluding the [SII] doublet from
the fit results in $\sim 35\%$ larger (NGC 3998) or smaller (NGC 4278)
dispersion. Similarly, Ferrarese \& Ford (1999) report for
NGC 6251 a width of the [SII] lines which is $ 37\%$ larger than the
{\HalphaNII} lines at the nuclear aperture. For the archival cases
with only narrow lines, we determined that the inferred gas
dispersion changes by $\lta 10\%$ by excluding the [SII] doublet from
the fit.

The next step is to find for each galaxy the aperture
closest to the nucleus which we will refer to as the 'central'
aperture. We assume that the dynamical nucleus coincides with the peak
in emission-line and continuum flux. The typical dimension of the STIS
apertures is $0.1''-0.2''$ in the spectral direction and
$0.05''-0.1''$ in the spatial direction (see Table~\ref{t:data}). The
exact sub-pixel location along the slit of the nucleus was
determined by fitting a Gaussian to the central few emission-line
fluxes. In most cases the peak in velocity dispersion coincides with
the flux peak. We indicate in Table~\ref{t:sample} the few cases in
which the maximum dispersion does not peak at the central aperture but
just outside it. The exact location of the nucleus (i.e., the flux
peak under our assumptions) in the direction of the slit width is not
known. We assume that it is centrally located in this direction. This
seems plausible given that the target acquisition procedure is
designed to have the center of the slit located exactly on the
brightest continuum point in the galaxy. That this 'peak-up' procedure
actually succeeded is supported by the steep decline in flux in
parallel slits which are available for many targets. Nevertheless, the
effect on our results of a potential off-set from this location is
discussed in Section~\ref{s:cave}. The gas velocity
dispersion $\sigma_{\rm g}$ at the central aperture is listed in
Table~\ref{t:sample}.  The typical relative formal measurement error
for $\sigma_{\rm g}$ is $\sim 5\%$.

We also require the unconvolved flux profile for our analysis. The
emission-line surface brightness is represented by fitting a double
exponential function,
\begin{equation}
  I(R) = I_1 \exp(-R/R_1) + I_2 \exp(-R/R_2),
\label{e:fluxparam}
\end{equation}
to the narrow emission-line fluxes taking into account the PSF and
disk inclination (see e.g., Verdoes Kleijn et al.~2002 for a detailed
description). The disk inclinations were taken from the
literature or determined by us from the extended dust disk often
present on HST imaging. If dust is absent or the inclination
could not be determined reliably, we assumed
$i=60\deg$.  The HST observational characteristics, unconvolved flux
profiles and disk inclinations are listed in Table~\ref{t:data}. The
fluxes are typically fitted within the formal errors or within 10\%. A
few galaxies have some asymmetry in their profiles over which the
symmetric fit averages, namely IC 1459, NGC 3245, NGC 4278,
NGC 4459 and NGC 4486.

Finally, we use published stellar dispersions (see
Table~\ref{t:sample}). The stellar dispersions $\sigma_{\rm s}$ are
measured through an aperture of typically several arcsec$^2$. This
corresponds to a region much larger than the BH sphere of influence,
and measures primarily the virial dispersion of the system as a
whole. We will make use of the stellar dispersions to estimate black
hole masses using the $\Mbh - \sigma$ relation as determined
by Tremaine \etal (2002). The relation is calibrated to flux-weighted
stellar dispersions $\sigma_{\rm e}$ inside an effective radius of the
galaxy, which are not available for most of our sample
galaxies. However, comparing $\sigma_{\rm e}$ and $\sigma_{\rm s}$
where possible shows a difference of typically $\sim 7\%$ between
$\sigma_{\rm s}$ and $\sigma_{\rm e}$. We will use this as the typical
relative error on our measurement of $\sigma_{\rm s}$ in approximation
to $\sigma_{\rm e}$. Lastly, Table~\ref{t:sample} also lists the black
hole masses obtained from dynamical modeling of central gas disk
rotation velocities as published for eight galaxies in the
sample.

\section{Thin Disk Models}
\label{s:excess}

Figure~\ref{f:sigmagassigmastar} shows the central gas velocity
dispersion at small scales (i.e., typical aperture size $\sim 0.15''$
or $\sim 40\pc$ for typical galaxy distance of $50\Mpc$) versus the
large-scale stellar dispersions (tracing the flux-weighted dispersion
at kpc scales).  The nuclear gas dispersions are almost always larger
than the large-scale stellar dispersions, increasingly so for larger
stellar dispersions. Given the success rate of black hole mass
determinations in early-type galaxies, it now seems quite possible
that all spheroids harbor a black hole at their nucleus. There are two
generic ways in which the presence of a BH can contribute to the
nuclear gas dispersion. First, gas motions increase at smaller
distances from a central black hole due to the increase of the
gravitational force. Integrated over a finite aperture this will lead
to a larger observed gas dispersion. We will call this the
gravitational contribution. An increase will occur always, but its
magnitude will depend on the physical structure and orbital
distribution of the gas, e.g., a thin rotating disk or a more
spheroidal distribution. The effect is noticeable because for a
typical $\sigma_{\rm s} = 250\kms$ in our sample, the typical central
HST aperture is only a factor two larger than the rough measure of the
black hole radius of influence $r_{\rm BH} = G \Mbh / \sigma_{\rm s}^2
\sim 20\pc$ adopting the $\Mbh - \sigma$ relation from
Tremaine \etal (2002). The relation finds $\Mbh \sim \sigma_{\rm s}^\alpha$ with $\alpha \sim 4$. If the gas motions are dominated by gravitation it is expected that $\Mbh \sim \sigma_{\rm g}^\beta$ with $\beta \sim 2$ and hence $\sigma_{\rm g} \sim \sigma_{\rm s}^2$ qualitatively consistent with the increasing ratio of $\sigma_{\rm g}$ and $\sigma_{\rm s}$ shown in Figure~\ref{f:sigmagassigmastar}. 
The $\Mbh - \sigma$ relation also suggests that one should
expect a smaller excess gas dispersion in galaxies with smaller
$\sigma_{\rm s}$ for another reason: galaxies with small $\sigma_{\rm s}$ 
have a smaller BH mass
and a sphere of influence so that a STIS aperture of fixed size
samples more of the gas outside of the BH sphere of
influence. A second generic way in which the presence of a BH can contribute to
the nuclear gas dispersion is not through its gravitational force, but 
through input of kinetic energy which perturbs the collisional gas. 
We will call this the
hydrodynamical or non-gravitational contribution. This energy could be released by
processes related to an active black hole. As a side remark, it is
unlikely that the kinetic energy might be provided by collisions with
photons emerging from the active region around the black hole. The
reason is that the bolometric luminosity of the low-luminosity active nuclei discussed here is
always orders of magnitudes below the Eddington luminosity (e.g., Ho
1999). Shocks, e.g., jet-gas interactions are a more plausible source
of kinetic energy. It could be that the gravitational contribution
dominates (e.g., a thin disk of non-colliding gas particles in
circular rotation) or that the hydrodynamical contribution dominates
(e.g., a fully collisionally driven outflow of gas from the active
black hole). The well-defined thin dust and gas
disks often seen immediately outside the nucleus in these galaxies
suggest that the gas might be settled all the way to the
nucleus. Therefore, we will first determine the answer to the
question: in which galaxies can the BH potential account for the
observed nuclear gas dispersion assuming that the gas is located in a
thin circular disk?

The answer to this question is known already for five
galaxies in our sample for which detailed modeling has been performed
of both gas velocities and dispersions of the extended gas disk under
the assumption of a thin circular disk. In the case of NGC 7052 (van
der Marel \& van den Bosch 1998), IC 1459 (Cappellari \etal 2002;
Verdoes Kleijn \etal 1999) and NGC 4335 (Verdoes Kleijn \etal 2002),
an excess of dispersion by a factor $\sim 2$ or more was observed. For
NGC 3245 (Barth \etal 2001), a smaller excess of $\sim 35\%$ was
observed. In the case of M87 (Macchetto \etal 1997; Harms \etal 1994),
no dispersion excess was observed. The velocity dispersion for
this galaxy was accounted for by a model in which the gas resides
in an annulus instead of a disk. (This model can also account well for
the velocities and peculiarities in the flux distribution.) It is
these mixed results on excess gas dispersion that
lead us to examine the presence of excess gas dispersion for a larger
sample of galaxies.

In the present and following sections
(Section~\ref{s:excess}--\ref{s:nongrav}) we present results of
dynamical models of the gas velocity dispersions. Caveats that
result from the assumptions in our models are discussed in
Section~\ref{s:cave}. Performing gas disk kinematical modeling for our complete sample in
similar detail as done for the aforementioned few cases is beyond the
scope of this paper. Thus we constructed somewhat more
simplified thin disk models. The main simplification is to neglect the
stellar mass contribution to the central gravitational
potential. The effect of this simplification on the predicted
velocity dispersions for the galaxy sample can be estimated as
follows. Assume a spherical stellar mass density $\rho(r) =
\rho_0 (r/r_0)^{\alpha}$ in the nuclear region, where $r$ is the radius in
pc and $\rho_0$ a scaling constant. According to the LEDA database, the absolute
blue magnitude in our sample varies in the range $M_B=[-19.3,-22.5]$
with a mean $<M_B>=-21.1$. Assuming a typical galaxy color $B-V=0.95$,
this indicates that a typical galaxy in our sample is about 1
magnitude brighter than the division magnitude at $M_V \sim -21.0$
between shallow core and steep core (``power-law''))
galaxies (Gebhardt \etal 1996). Galaxies brighter than this division
magnitude typically have $\alpha \sim -1$ (while below
they have $\alpha \sim -2$).  From the study by Gebhardt \etal (1996)
we infer a typical $\rho_0 \lta 10^4 \Msun/{\rm pc}^{3}$. With
this information we can compute the circular velocity $v_{c,{\rm s}}$
due to the stellar mass at a given radius as $v_{c,\rm{s}}^2=4\pi G
\rho_0 \frac{r^{\alpha+2}}{\alpha+3}$. By comparison, the 
circular velocity due to the BH is $v_{c,\rm BH}^2 = \frac{G
\Mbh}{r}$.  We compute the ratio of these two at $r=0.1''$ from the
nucleus assuming a BH mass according to the $\Mbh - \sigma$
relation. The value $r=0.1''$ is a typical distance
from the galaxy center at which we have measured gas dispersions with
HST. The ratio of the circular velocities in quadrature (i.e.,
equivalent to the ratio in dynamical mass) is always less than 20\%.
This is much smaller than, e.g., the scatter of $\sim 0.3$ dex in the
$\Mbh - \sigma$ relation (Tremaine \etal 2002). This confirms
that the stellar mass contribution in our analysis can be safely
neglected, provided that: (a) the BH masses are not significantly
below the $\Mbh - \sigma$ relation; and (b) we are not
attempting to model the large scale rotation of the gas at radii much
beyond $\sim 0.1''$. As an additional check we directly
determined the difference in predicted dispersion with and
without BH mass for NGC 4335, NGC 7052 and IC 1459. For these
galaxies, we modeled the stellar mass in previous papers (Verdoes
Kleijn \etal 2002; van der Marel \& van den Bosch 1998; Verdoes Kleijn
\etal 1999). These cases confirm the result from the general argument
that the difference in predicted dynamical masses is always less
than 20\% (see also Figure~\ref{f:mbhlit}, discussed
below).

For a thin circular disk, the velocity dispersion observed in a
nuclear aperture is due entirely to differential rotation over the
aperture. To calculate this dispersion we used the
modeling software that is described in more detail in van der Marel \&
van den Bosch (1998) and Verdoes Kleijn \etal~(2000; 2002). This
modeling takes into account the emission-line flux profile, slit
width and PSF convolution. We first modeled
the galaxies for which a BH mass estimate is available in the
literature from detailed modeling of the gas rotation
velocities. Figure~\ref{f:mbhlit} shows the predicted versus observed
velocity dispersions for these eight galaxies. Seven
galaxies have an observed dispersion in excess of the prediction. The
modeling nicely reproduces the aforementioned relative
differences in excess dispersion for M87, NGC 3245, NGC 7052, NGC 4335
and IC 1459 known from prior modeling of the
rotation velocities in the extended gas disk.
To put the complete sample in a similar diagram, we need an
estimate of the BH mass for every galaxy. For this we use
the $\Mbh-\sigma$ relation as calibrated by Tremaine
\etal (2002). They derive the relation $\log [\Mbh (\Msun)] = 8.13 +
4.02 \log (\sigma/ 200 \kms)$. The slope of this relation
falls between those reported by Gebhardt \etal (2000; slope=3.75) and
Ferrarese \& Merritt (2000; slope=5.27). Figure~\ref{f:mbhdisk}
shows the predicted versus observed dispersion for a thin
disk model, but now assuming for {\it all} galaxies a
BH mass according to the $\Mbh - \sigma$ relation.  The observed dispersions in non-radio galaxies scatter around the prediction from the model (except for IC 989: cf.\ Section~\ref{s:conc}). In fact, the number of non-radio galaxies and their range in observed gas dispersions is large enough that they independently confirm the $\Mbh-\sigma$ relation for the thin disk model. This is non-trivial because BH masses significantly below the $\Mbh-\sigma$ relation could have been detected with the HST/STIS observations. As a measure of the minimally detectable BH mass, we compute the BH mass which causes a gas dispersion which equals that predicted from the combined effect of instrumental broadening and the stellar mass model discussed above. We use again the typical radius of $0.1''$ and take into account the disk inclination and distance for each galaxy. The resulting masses vary between $\sim 6 \times 10^6\Msun$ and $ \sim 6 \times 10^7 \Msun$ with a typical value of $\sim 3 \times 10^7\Msun$. (Exceptions are NGC 3862 and UGC 7115. Their minimally detectable BH mass lies above the $\Mbh-\sigma$ relation because the instrumental line broadening is larger than the dispersion due to differential gravitational rotation for such close to face-on gas disks.) The BH masses predicted by the $\Mbh-\sigma$ relation exceed this typical detectable BH mass by a factor $\sim 2$ for the lowest stellar dispersions and by almost 2 orders of magnitude for the highest stellar dispersions in the sample. In conclusion, the gas dispersions in non-radio galaxies support the idea that all early-type galaxies harbor supermassive black holes at their nuclei with masses according to the $\Mbh-\sigma$ relation.

In contrast to the non-radio galaxies, all radio galaxies have a gas
dispersion in excess of that expected from the thin disk models. About
$\sim 50\%$ of those fall above the $1\sigma$ scatter expected from
the intrinsic scatter in the $\Mbh - \sigma$ relation. To explore the
origin of this we plot in Figure~\ref{f:sigratiodisk} the ratio
$R_{\rm o,d}$ of observed dispersion and that predicted for the disk
model. There is a trend of an increasing ratio with decreasing disk
inclination. In fact, the four largest ratios are all in galaxies with
gas disk inclinations $i<40\deg$, i.e., close to face-on disks. This
cannot be explained as a result of measurement errors in the dust disk
inclinations. The arrows in Figure~\ref{f:sigratiodisk} show by how
much $R_{\rm o,d}$ can decrease due to the known measurements errors.
The decreases do not change the overall result that radio galaxies
systematically have excess dispersions. A more extreme (and probably
unphysical) possibility is to assume that we might have used incorrect
inclinations because the measured dust disk inclinations bear no
relation to the actual inner gas disk inclinations. However, this
would still not provide a satisfactory explanation. If all the gas
disks were in reality edge-on and well described by thin disk models,
then we would have expected the points in Figure~\ref{f:sigratiodisk}
to fall along the curve $R_{\rm o,d} = \sin^{-1} i$ (solid
curve). This still falls below many of the observed dispersion ratios
(see Figure~\ref{f:sigratiodisk}). Moreover, an excess gas dispersion
is not only seen in radio galaxies with relatively face-on disks. If
we exlude systems with disk inclinations $i<40\deg$ from the analysis,
the median $R_{\rm o,d} = 1.32$ for radio galaxies versus $0.96$ for
non-radio galaxies. A median test yields a probability of only 0.05
that this difference would occur by chance in the event that the
samples were drawn from populations with the same median. Therefore, a
solution to account for the excess gas dispersion in radio galaxies
must probably be sought in having a significant vertical velocity
dispersion component.

Finally, as a sanity check we
analyzed the dependence of the $R_{\rm o,d}$ on BH mass and the
typical circular velocity of the system, parameterized as
$\sqrt{\frac{G \Mbh}{r_{\rm HALF}}}$ where $r_{\rm HALF}$ is the
radius inside which half of the light of the unconvolved emission-line
flux distribution is contained. There is no relation between the
dispersion ratio and the two measures of the gravitational potential.
This confirms that the excess velocity dispersion in the
radio galaxies is not somehow an artifact of our gravitational modeling.

\section{Spheroidal Models}
\label{s:spheroid}

Many galaxies in the sample show disks of dust and gas with radii
ranging from hundreds of pc up to several kpcs. It appears reasonable,
but might be wrong to assume by extrapolation that the
nuclear region within the spatial resolution of the observations
also has a disk-like geometry. Instead, the gas
distribution might be more spheroidal in the vicinity of the black
hole. In that case the gas moves also out of the plane of the larger
scale thin gas disk. This could explain qualitatively the
observed increasing ratio of observed versus predicted dispersion for
decreasing dust disk inclination. We address next whether
such a spheroidal distribution can also quantitatively
explain the observed $\sigma_{\rm g}$ gravitationally. 

Gas in a vertically extended distribution (e.g., a spherical one)
tends to collapse quickly to a disk due to hydrodynamical
forces. Thus, to maintain a spheroidal structure, one would have to
assume that the gas is located in individual clouds that move
collisionlessly. As an initial assessment of the plausibility of this
we estimate the collision time-scale for clouds in a simplistic model:
spherical cloudlets of ionized hydrogen moving around in a spherical
volume. The typical time $t_{\rm free}$ between cloudlet collisions is
estimated as:
\begin{equation}
t_{\rm free}=\frac{V}{N_{\rm cl}*\sigma_{\rm cl}} \frac{1}{v_{\rm cl}},
\end{equation} 
where $v_{\rm cl}$ is the typical relative velocity of cloudlets, $V$
is the spherical volume within which the cloudlets are contained,
$\sigma_{\rm cl}$ is the cross-section of each cloudlet and $N_{\rm
cl}$ is the total number of clouds. For $V$ we take the spherical
volume with a radius equal to the half-light radius of the gas
distribution as seen projected on the sky. For $v_{\rm cl}$ we take
the circular velocity at half the half-light radius. The number of
cloudlets $N_{\rm cl}$ is computed from
\begin{equation}
N_{\rm cl} = N / ( \frac{4}{3}\pi R_{\rm cl}^3 n_e),
\end{equation}
where $N$ is the total number of electrons, $n_{\rm e}$ is the
electron density and $R_{\rm cl}$ is the radius of each cloudlet. The
electron density can be obtained from the relation between the flux
ratio [SII]$\lambda$6716/$\lambda$6731 and the electron density
(Osterbrock 1989). This flux ratio is only sensitive to electron
densities in the range $[10^2,10^4]$cm$^{-3}$. Outside this range the
flux ratio becomes constant as a function of $n_e$. We obtain the flux
ratio for 15 of the radio galaxies from Noel-Storr \etal (2003) of
which six lead to upper and lower-limits on the electron density. $N$
can be estimated as discussed in Osterbrock (1989) assuming case B
recombination:
\begin{equation}
N = \frac{L_{H\beta}}{n_e \alpha_{H\beta}^{eff} h \nu_{H\beta}},
\end{equation}
where $\alpha_{H\beta}^{eff}=3.03\times 10^{-14}$cm$^{-3}$ is the
recombination coefficient (assuming $T=10^4$K) and $\nu_{H\beta}$ and
$L_{\rm H\beta}$ are the frequency and luminosity for H$\beta$.  We
obtain $L_{H\beta}$ from the central H$\alpha$+[NII] luminosity
(Verdoes Kleijn \etal~2002) using the standard
$L_{H\alpha}/L_{H\beta}$ ratio of 3.1 and the
[NII]$\lambda\lambda$6548,6584 / H$\alpha$ flux ratio determined from
our Gaussian line fits.  This results in the following equation for
the collision time:
\begin{equation}
t_{\rm free} \sim \frac{R^{3.5} n_e^2 R_{\rm cl}}{\sqrt{\Mbh} L_{H\beta}}  
\end{equation}
This increases linearly with the radius of the cloudlets $R_{\rm cl}$.
If we assume that $R_{\rm cl}$ is as large as a tenth of the
half-light radius, then the inferred $t_{\rm free}$ varies between
$10^3$yr and $10^{13}$yr with a typical value of $\sim
10^{7-8}$yr. Some of the huge scatter is most likely due to the
simplicity of the estimate. But either way, it appears unlikely that
the gas could remain in a collisionless state for a significant
fraction of the Hubble time. Collisions lead to dissipation, loss of
energy, and settling onto circular orbits.  However, this only true if
there is no energy input into the gas.  That may well be incorrect for
most of the galaxies in our sample, given that 16 galaxies are FR I
radio galaxies, and the remainder have nuclear radio emission. If
there is energy input, then the gas can maintain an extended
distribution for much longer than the timescale $t_{\rm free}$
calculated above. Even though collisions would occur, one might expect
the system to evolve through a sequence of states that are all
approximately collisionless. Therefore, it is reasonable to study the
predictions of collisionless models for the nuclear gas in our sample
galaxies.

A major uncertainty in the calculation of collisionless models is the
phase-space distribution of the clouds, which is needed to compute the observed
velocity dispersion along the line of sight. The phase-space
distribution involves a spatial component such as a radial number
density profile.  We do not know this profile. We only know the flux
distribution, which could be quite different. The phase-space
distribution also requires knowledge about the nature of the cloud
kinematics, for example anisotropies in the velocity dispersion.
These are also unknown. We therefore decided to explore the line-of-sight velocity
dispersion for a plausible range of phase-space distributions.  We use
axisymmetric dynamical models presented in de Bruijne \etal (1996). We
explore gas clouds with an axisymmetric power-law number density
distribution with axis ratio $q$ (i.e., $\rho(R,z)=\rho_0
[R^2+z^2/q^2)^{(-\gamma/2)}]$) in the Kepler potential caused by
the black hole. The gas clouds are assumed to have a constant velocity
dispersion anisotropy $\beta=1-(\overline{v_\theta^2}+
\overline{v_\phi^2}) / (2 \overline{v_r^2})$ 
(i.e., the 'case II' models in de Bruijne \etal 1996; cf.~Binney 1980,
with $(r,\theta,\phi)$ the usual spherical coordinates). Thus we have
four free parameters: the number density scaling constant $\rho_0$,
the axis ratio $q$, the power-law slope $\gamma$ of the particle
density profile and the velocity dispersion anisotropy
$\beta$. These models require $\gamma>1/2+\beta$ to be physical. De
Bruijne \etal (1995) evaluated the projected velocity
dispersions. The software to calculate them numerically is
available from http://www.stsci.edu/{$\sim$}marel/.

We want to know how different the observed velocity
dispersion for a spheroid can be compared to the dispersion
for a thin disk, for a plausible range of the free
parameters. Thus we determined the ratio $R_{\rm s,d}$
of the velocity dispersions for a spheroid and a thin disk
that have the same projected surface density
distribution and apparent axial ratio. The latter criterion fixes
the ratio of the $\rho_0$ values for the two models, and implies that
$R_{\rm s,d}$ is independent of $\rho_0$. We define $R_{\rm s,d}$ 
as the ratio of the line-of-sight dispersion weighted by the number
density and integrated over an aperture which is typical for the
observations. The result depends on power-law slope, axis
ratio and dispersion anisotropy. We explored the
parameter ranges $\gamma=[-1.5,-10]$, $q=[0.1,0.9]$ and
$\beta=[-\inf,1]$, which should encompass all plausible
models. Figure~\ref{f:spheroid} shows $R_{\rm s,d}$ as a
function of apparent axis ratio for the quoted parameters
ranges. Regardless of the choice of parameters, the modeled spheroidal
distributions have increased velocity dispersions in comparison
to thin disks only for $q \gta 0.7$ (corresponding to
inclinations $i \lta 45\deg$ for a thin disk).

As also shown in Figure~\ref{f:sigratiodisk}, the non-radio
galaxies scatter around $R_{\rm o,d} = 1$ indicating that their
dispersions are consistent with thin disks in circular rotation
around BHs that follow the $\Mbh - \sigma$ relation.  By
contrast, the radio galaxies have dispersion ratios $R_{\rm o,d}$ which are
systematically larger than 1, and also
larger than the $R_{\rm s,d}$ expected for
collisionless spheroidal models. So it seems
unlikely that a spheroidal distribution of collisionless gas
clouds can account for the nuclear gas velocity dispersions in
radio galaxies. We also explored the ratio of the
velocity dispersions for an axisymmetric spheroid and a disk
at the same inclination instead of the same apparent axis
ratio\footnote{Note that in this case that the disk and spheroid do
not have an identical surface density distribution.}. The reason is
that the axis ratio of the gas distribution at the nucleus is not
constrained very well by the observations and might deviate from the
axis ratio of the larger scale disk. The resulting $R_{\rm s,d}$ range for spheroid models at a given
inclination remains very similar. So this does not alter the
conclusions inferred from Figures~\ref{f:sigratiodisk} and~\ref{f:spheroid}.

\section{The non-gravitational gas dispersion component}
\label{s:nongrav}

The analysis of thin disks and spheroidal models indicates
that non-gravitational forces might contribute significantly to the
observed gas dispersion for radio galaxies. To constrain the minimal
relative contribution of a non-gravitational dispersion component we
plot in the top panel of Figure~\ref{f:nongrav} the ratio of the
observed dispersion and the maximum dispersion predicted by
either disk or spheroid models. The minimal ratio
is typically between 1 and 2. There is a hint that the minimal ratio
increases with axial ratio, i.e., for extended disks that are
closer to face-on. This could indicate that the non-gravitational
dispersion component does not have a random orientation with respect
to the extended disk but is oriented preferentially along the
spin axis of the extended disk. We explore this idea with
two simple models. The models consist of two components: a thin
circular rotating disk and a second component of non-gravitational
motions. We assume equal contribution to the emission-light by the two
components. In the 'isotropic model' the non-gravitational dispersion
is isotropic ($\sigma_{\rm iso}$) and the resulting total dispersion
can be written as
\begin{equation}
\sigma^2_{\rm obs} = \sigma^2_{\rm grav}\sin^2 i +\sigma^2_{\rm iso}/3,
\label{e:sigmanongraviso}
\end{equation}
where $\sigma_{\rm grav}$ denotes the dispersion due to circular rotation as measured in the plane of the disk and $i$ is the disk inclination. In the 'perpendicular model' the non-gravitational dispersion is assumed to be perpendicular to the plane of the disk:
\begin{equation}
\sigma^2_{\rm obs} = \sigma^2_{\rm grav}\sin^2 i +\sigma^2_{\rm perp}\cos^2 i.
\label{e:sigmanongravperp}
\end{equation}
The middle and bottom panels of Figure~\ref{f:nongrav} show the
ratios $\sigma_{\rm iso}/\sigma_{\rm grav}$ and $\sigma_{\rm
perp}/\sigma_{\rm grav}$ as a function of disk axis
ratio. The value of $\sigma_{\rm grav}$ is calculated as before
from the thin disk models and $\sigma_{\rm iso}$ and $\sigma_{\rm
perp}$, respectively, are chosen so that $\sigma_{\rm obs}$ matches the
observed value $\sigma_{\rm g}$. The perpendicular model shows no
trend between the dispersion ratio and disk inclination. The isotropic
model shows a weak trend. This would not be expected in an
isotropic scenario. On the other hand, the trend in
Figure~\ref{f:nongrav}b is dominated by a few galaxies and is not
really significant. For some galaxies the implied
non-gravitational dispersion is larger than $\sqrt{2}\sigma_{\rm
grav}$. In other words, the typical non-gravitational motions are
larger than $\sqrt{2}$ times the typical circular rotation in those
galaxies which suggests that some of the material might be
unbound. However, the models are too simplistic to
attach real significance to this result.

The fact that radio galaxies show an excess gas dispersion, while it
appears to be absent in non-radio galaxies, possibly points to a link
between non-gravitational motion and activity. The AGN could be
a source of energy to drive the non-gravitational dispersion e.g., by
coherent flows or turbulence. We do not have a direct measure of the
kinetic energy output of the AGN to determine its correlation with the
amount of non-gravitational dispersion present in the gas. Thus we
resort to measures of the radiative energy output of the active nuclei
as a proxy.  Figure~\ref{f:agn} shows $R_{\rm o,d}$ as a function of
total radio power for both radio and non-radio galaxies. As expected,
the difference in excess dispersion for non-radio and radio galaxies
is also present as a function of radio luminosity. However, there is
no clear indication that larger non-gravitational motions occur in
more powerful radio galaxies. We examined also the dependence of
$R_{\rm o,d}$ on radio core and nuclear emission-line luminosity for
radio galaxies. No clear trends were found. Similarly, no
trends were found between the ratios $\sigma_{\rm
iso}/\sigma_{\rm grav}$ and $\sigma_{\rm perp}/\sigma_{\rm grav}$
with these three indicators of the radiative power of the active
nucleus.

\section{Caveats}
\label{s:cave}

{\bf Uncertainties in the flux profiles} For three
galaxies in our sample, NGC 3862, NGC 4374 and NGC 4486, we have both
the STIS flux profiles and those derived from HST/WFPC2 PC-chip
emission-line images which are available to us. The PC pixel size
($\sim 0.045''$) better samples the HST PSF FWHM
($\sim 0.1''$) compared to the STIS
apertures. From the PC emission-line image, a narrower flux
profile is inferred for NGC 3862 and NGC 4374, but a wider profile for
NGC 4486, as compared to the flux profiles derived from the STIS
data. Using  the PC profiles in the thin disk modeling
leads to increases in predicted gas velocity dispersion of $\sim 5\%$,
$\sim 50\%$ and $\sim 13\%$ for NGC 3862, NGC 4374 and NGC 4486,
respectively. Assuming that similar changes could be expected for the
other sample galaxies, this does not change our conclusions.

Another
possibility to consider is that we might have
systematically overestimated the width of the emission-line flux profiles. If
the flux in reality originates closer to the nucleus then in our
models, it would naturally explain why the predicted dispersions are
lower than observed. Figure~\ref{f:halflight} shows the radius
inside which half of the flux from the intrinsic (i.e., unconvolved)
flux profile is contained as a function of distance. The top panel
shows the angular half light radius. For half light radii much less
than the half light radius of the HST PSF, i.e., $\sim 0.05''$, it
could well be that the intrinsic flux profile is in fact much
narrower. Thus, the profile widths for NGC 2329, NGC 3862 and NGC 6251
should be considered upper limits. This implies that their predicted
velocity dispersions should be considered lower limits as
emission-line gas might be rotating closer to the black hole than
in our models. However,  there are two facts that argue against a
general overestimate of the width of the flux profiles
for all galaxies. First, we find no trend of narrower
emission-line flux profiles with increasing dispersion ratios. Second,
such a scenario cannot explain the correlation between disk
inclination and dispersion ratio. A separate question is whether
we are resolving the flux profile at larger distances as well as
nearby. This issue is relevant for comparing the properties
of the radio galaxies and non-radio galaxies, since the
former sample has a larger average distance.  The bottom panel of
Figure~\ref{f:halflight} suggests that this distance offset is not
inducing the difference in excess dispersion, since there is no
trend between the physical half light radius and distance.

Finally, what kind of emission-line flux profile is needed to obtain the observed gas velocity dispersions gravitationally? To first approximation $\sigma_g^2 \sim \Mbh / R_{\rm flux}$, where $R_{\rm flux}$ is the typical radius of the flux profile (e.g., the half light radius). Figure~\ref{f:sigratiodisk} shows that for radio galaxies $R_{\rm flux}$ needs to decrease by a factor of about 2 up to 100 to have a model with a black hole mass according to the $\Mbh-\sigma$ relation account for the observed gas dispersion. Figure~\ref{f:halflight} shows that the emission-line flux profile for radio galaxies typically has a half light radius which is 2 times the radius of the PSF. In conclusion, the emission-line profile for radio galaxies needs to be fully unresolved in general to account for the observed gas velocity dispersion gravitationally. This is ruled out by the observations.

{\bf Uncertainties in the location of the black hole} The spatial flux
distribution in the direction of the slit length allows us to
precisely locate the nucleus along this direction (assuming it
coincides with the peak of the emission-line flux,
cf.~Section~\ref{s:sampledata}). But the nucleus was
{\it assumed} to be at the center of the slit in the
direction of the slit width. This is a reasonable assumption (see
Section~\ref{s:sampledata}). We determined that a $0.05''$
displacement in this direction changes the predicted velocity
dispersion by $<20\%$ for the thin disk models. This
does not affect our conclusions.

{\bf Does the $\Mbh - \sigma$ relation underestimate BH
masses?}  One might account for the excess gas dispersion within the
purely gravitational models, seen particularly for radio galaxies, 
by assuming that the BH mass from the
$\Mbh - \sigma$ relation underestimates the true BH
mass. However, this requires increases in BH masses that are
too large to be credible for the following reasons. First,
comparison of Figure~\ref{f:mbhlit} and
\ref{f:mbhdisk} shows that detailed estimates of BH masses
from gas rotation velocity modeling do not show any
indication for BH masses that are systematically larger than
those inferred from the $\Mbh - \sigma$ relation. (This is a
somewhat circular argument as some of the BH mass measurements from
gas were used to determine the $\Mbh - \sigma$
relation. Nevertheless, such large BH masses would imply velocity gradients which are larger than observed.) Second, the trend of an increasing $R_{\rm o,d}$
with decreasing disk inclination is not expected if the excess dispersion is due to underestimated BH masses. The dependence of BH mass on stellar dispersion is sometimes claimed to be steeper than used here (e.g., Ferrarese 2002). However, this amounts to at most $\sim 50\%$ larger BH masses and such uncertainties will not affect our results on excess gas dispersions. Lastly, it has been argued that $\Mbh$ correlates equally well with host luminosity as with $\sigma$ (e.g., Marconi \& Hunt 2002). This might mean that the $\Mbh-\sigma$ relation underestimates $\Mbh$ at the high $\sigma$ end. The reason is that Brightest Cluster Galaxies (BCGs) fall in this region. The BCGs are exceedingly luminous for their stellar dispersion which is typically above $270\kms$ (Fisher, Illingworth \& Franx 1995). If true, it is very unlikely that this will affect our results because the galaxies in our sample with the largest excess in stellar dispersion compared to gravitational models all have stellar dispersions between $190\kms$ and $270\kms$ (cf., Figure~\ref{f:magsigma}).

{\bf Are thin disk models plausible?} It was found that for most of
the normal galaxies in our sample the nuclear gas dispersions are
reasonably well fit by thin disk models containing BHs that follow the
$\Mbh - \sigma$ relation. But the full emission lines shapes
contain additional information on the line of sight velocity
distribution (LOSVD) and can further constrain the plausibility of the
models. We used the thin disk models to calculate the full LOSVD for
the normal galaxies in the sample. Inspection of the data shows a qualitative agreement between
   predicted and observed LOSVDs. A general property of the LOSVD of
   thin disk models is a tendency for being double-peaked. In some
   cases the predicted double peakedness seems slightly too pronounced
   to be consistent with the data. But it is possible that the
   predicted peaks have in reality been broadened away by additional
   velocity dispersion contributions. For example, the gas could be
   collisionless but orbit in a somewhat more three-dimensional
   distribution than a disk (as in the models of Section~\ref{s:spheroid}), or it
   could have a small additional non-gravitational contribution (as in
   the models of Section~\ref{s:nongrav}). Overall, the signal-to-noise ratio of the
   data and the blending of the H +[NII] lines make it difficult to
   draw strong conclusions from the observed line shapes.

{\bf Could an unaccounted broad H$\alpha$ component in radio
galaxies influence our results?} The flux from which the velocity
dispersions are measured is usually dominated by the flux from the
forbidden [NII] and [SII] lines, especially the [NII]$\lambda6584$ line
and not the H$\alpha$ line. Furthermore, we ensured to correct for
the presence of broad H$\alpha$ if it was clearly detected. This broad
component is likely arising from a broad-line region which is located
much closer to the nucleus than the narrow component that we model
here. Noel-Storr \etal (2003) found tentative indications that a broad
line, not readily seen by eye, could be present in eight of the radio
galaxies in our sample (Table~\ref{t:sample}). In these cases they
performed fits to the H$\alpha$+[NII],[SII] complex assuming the
permitted H$\alpha$ line has both a narrow and a broad component. We
redid the thin disk modeling for these eight galaxies taking into
account the tentative broad component in measuring the velocity
dispersion and emission-line profile of the narrow component. This
resulted in decreases of $R_{\rm o,d}$ by at most $20\%$ and a median
decrease of $9\%$. Thus these eight galaxies do not change the overall
results obtained together with the 13 other radio galaxies.
Lastly, the dispersion in the extended disk are in many radio galaxies above those expected from instrumental effects and differential rotation over the aperture (e.g., Noel-Storr \etal 2003). This strengthens the idea that the excess dispersions are not due to a nuclear broad line region. 

{\bf Gaussian fit versus true second velocity moment.} For the thin
disk models we compared the predicted second moment of a Gaussian fit
to the full line of sight velocity distribution (LOSVD) to the
observed Gaussian fit. By
contrast, in the analysis of more spheroidal distributions,
$R_{\rm s,d}$ is the ratio of the true second
moments. Thus we assume that the ratio of the true second moments and
the ratio of the Gaussian-fit moments do not differ
significantly. It is not straightforward to test this assumption
quantitatively,  since the line profiles are not straightforward
to calculate for the spheroidal models. However, we believe that
the errors thus introduced are unlikely to exceed several tens of
percent (van der Marel \& Franx 1993). This is insufficient to affect
any of our conclusions. Also, it seems unlikely that a difference
in these ratios would affect our main result that there is a
systematic difference in nuclear dispersions between radio and
non-radio galaxies.

{\bf Dependence of the stellar dispersion on inclination.} In our
models we have assumed that the BH mass is uniquely determined by the
stellar velocity dispersion. However, this can only be approximately
true in reality. There must be some intrincic scatter in the
$\Mbh-\sigma$ relation, if only because the observed velocity
dispersion of a galaxy generally depends on the inclination under
which the galaxy is viewed, whereas the BH mass does not. This could
be important in the context of our results. The four galaxies for
which thin disks models fit the observed gas dispersions most poorly
have below-average inclinations, in addition to being radio galaxies
(see Figure~\ref{f:mbhdisk}). It is therefore important to study the
relation between between stellar velocity dispersion and inclination
in some more detail.

To support its shape, an oblate stellar system must on average have
more pressure ($\rho \sigma^2$) parallel to the equatorial plane
($\rho \sigma_{\parallel}^2 \equiv \rho [\overline{v_\phi^2} +
\overline{v_R^2}] / 2$) than perpendicular to it ($\rho \sigma_{\perp}^2
\equiv \rho \overline{v_z^2}$). Therefore, it will generally have a lower
line-of-sight velocity dispersion when viewed face-on than when viewed
edge-on. The tensor virial theorem (Binney \& Tremaine 1986) gives the
ratio of the pressures when integrated over the entire galaxy as
\begin{equation}
  (\sigma_{\parallel} / \sigma_{\perp})^2 = {1 \over 2} {
    { \Bigl ( {1 \over {\sqrt{1 - e^2}}}
                    {{\arcsin e} \over e} - 1 \Bigr ) } \over
    { \Bigl ( 1 - \sqrt{1 - e^2} {{\arcsin e} \over e}
      \Bigr ) } } , \quad\quad e^2 \equiv 1 -q^2_t .
\end{equation}
where $q_t$ is the intrinsic axial ratio. For
example, this gives $\sigma_{\parallel} / \sigma_{\perp} = 1$ for a
spherical galaxy and $\sigma_{\parallel} / \sigma_{\perp} = 1.34$ for
a galaxy with $q = 0.5$. If one were to observe a galaxy with an
aperture of infinite size, and if the galaxy had $\overline{v_\phi^2}
= \overline{v_R^2}$ at all positions in the galaxy, then the ratio of
the measured dispersion when viewed edge-on and face-on, respectively,
would be equal to $\sigma_{\parallel} / \sigma_{\perp}$. However,
these requirements are not generally met for real galaxies and real
observations. There is no reason why galaxies should have
$\overline{v_\phi^2}/ \overline{v_R^2} = 1$, and observational
constraints on this ratio remain scarce (e.g., Gebhardt et al.~2003).
Also, velocity dispersion measurements are generally restricted to the
central region of the galaxy and/or a specific axis (most often the
major axis) and do not integrate over the entire galaxy. So the true
dependence of the observed stellar velocity dispersion on the
inclination angle can only be addressed with detailed three-integral
stellar dynamical models (e.g., Gebhardt et al.~2003; Cappellari et
al.~2005b), which are outside the scope of the present
work. Nonetheless, the tensor virial theorem does provide useful
order-of-magnitude guidance. Almost all elliptical galaxies have axial
ratios $q > 0.5$ (Franx, Illingworth \& de Zeeuw 1991; Tremblay \&
Merritt 1995). Therefore, one might expect that inclination effects
cause a variation of order $\pm 15$\% ($\pm 0.06$ dex) in the observed
velocity dispersion of galaxies of otherwise identical properties.

An alternative way to constrain the inclination dependence of the
velocity dispersion is through the scatter in several well-know
relations. The velocity dispersions correlates strongly with galaxy
luminosity (the Faber-Jackson relation; e.g., Dressler et al.~1987),
with BH mass (the $\Mbh-\sigma$ relation), and with galaxy
mass-to-light ratio $M/L$ (Cappellari et al.~2005b). None of the latter
quantities depends on the viewing angle of the observations, while the
velocity dispersion does. Therefore, the observed scatter in these
relations sets an upper limit to the RMS variation that the
inclination might induce in the observed velocity dispersion. The
limits are $0.25/3.50 = 0.072$ dex from the Faber-Jackson relation (Dressler et al.~1987),
$0.30/4.02 = 0.075$ dex from the $\Mbh-\sigma$ relation (Tremaine et al.~2002), and
$0.07/0.82 = 0.083$ dex from the relation with $M/L$ (Cappellari et al.~2005b). These results
are all in reasonable agreement with the expectation from the virial
theorem (despite the fact that each of these relations is likely to
have other sources of intrinsic scatter as well). Lastly, direct estimates of the dispersion anisotropy in giant ellipticals from detailed three-integral
stellar dynamical models indicate a change of $\sim 0.05$ dex in stellar dispersion for face-on and edge-on viewing angles (Cappellari \etal 2005a). Given that rotational velocities contribute typically less than $10\%$ to the second moment of giant ellipticals this lends further support for the estimates based on global relations for the dependence of the stellar dispersion on inclination.

To obtain more specfic information for the actual galaxies in our
sample we plot their Faber-Jackson relation in
Figure~\ref{f:magsigma}. The four galaxies with dust/gas disk
inclinations $i<40\deg$ are indicated with special symbols. If the
disks reside in the equatorial plane of the galaxies, then the disk
inclinations are identical to the galaxy inclinations. One would then
expect the galaxies with $i<40\deg$ to have relatively low stellar
velocity dispersions for their luminosity. This is exactly what the
figure shows. The galaxies with $i<40\deg$ seem to lie $\sim 17$\%
($0.079$ dex) below the average relation between galaxy magnitude and
stellar dispersion. This agrees to lowest order with the expectation
from the tensor virial theorem and the scatter in well-known
correlations.

So there is theoretical and observational support for the idea that
the observed stellar velocity dispersion depends on the galaxy
inclination angle. By using the $\Mbh-\sigma$ relation to estimate the
BH mass one is therefore likely to underestimate the BH mass of
galaxies with relatively low inclinations. For the galaxies in our
sample with $i<40\deg$ we may have underestimated the BH mass by a
factor $0.83^{-4.02} = 2.1$ . The gas dispersions predicted by our
thin disk models will then be too low by a factor of $\sqrt{2.1} =
1.4$. Applying this correction to the predicted gas dispersions for
the $i<40\deg$ galaxies in e.g., Figure~\ref{f:mbhdisk} somewhat
reduces the discrepancy between the observed and predicted gas
dispersions for these galaxies. Nonetheless, for all four of the
galaxies the observed gas dispersions remain well in execss over those
predicted by the thin disk models. So our conclusion that radio
galaxies have a contribution from non-gravitational motions in their
central gas dynamics remains unaffected.

\section{Conclusions \& Discussion}
\label{s:conc}

We have analyzed HST/STIS observations of a sample of 27 galaxies.
The nuclear velocity dispersion of the gas in a STIS aperture of $\sim
0.1''$--$0.2''$ (i.e., scales of tens of parsecs) generally exceeds
the large-scale stellar velocity dispersion of the galaxy. This is
qualitatively consistent with the presence of central BHs, but raises
the questions whether the excess gas dispersion is of gravitational or
non-gravitational origin and whether the implied BH masses are
consistent with our current understanding of BH demography. To address
these issues we have constructed purely gravitational axisymmetric
dynamical models for the gas, both thin disk models and models with
more general axis ratios and velocity anisotropies. This has yielded
the following conclusions:

\begin{itemize}

\item{For the normal galaxies in the sample (i.e., without large-scale 
radio-jets) the nuclear gas dispersions are adequately reproduced by
models that have purely gravitational motion and BHs that follow the
$\Mbh-\sigma$ relation. Among the purely gravitational models
we cannot generally discriminate between thin disk models and vertically
extended models. The former might seem preferred theoretically because
they represent a longer-term equilibrium configuration for the
gas. However, in some cases it is unclear if the observed line profiles are consistent with the
double-peaked structure that is generally predicted by thin disk
models.}

\item{The nuclear gas dispersions observed for the radio galaxies generally 
exceed those predicted by models with only gravitational motions in
either a thin disk or a more spheroidal gas distribution. We attribute
this to the presence of non-gravitational motions in the gas that are
similar to or larger than the gravitational motions. The
non-gravitational dispersion is consistent with being either isotropic
or perpendicular to the extended gas and dust disks. The
non-gravitational motions are presumably driven by the active galactic
nucleus (AGN), but we do not find a relation between the radiative
output of the AGN and the non-gravitational dispersion.}

\item{Given the uncertainties about the dynamical state of the
gas, it is not possible to uniquely determine the BH mass for each
galaxy from its nuclear gas dispersion. However, for the sample as a
whole the observed dispersions do not provide evidence for significant
deviations from the $\Mbh-\sigma$ relation in either active or
non-active elliptical galaxies. In no case is the observed nuclear gas
dispersion so low that it puts an upper limit on the BH mass that is
significantly below the $\Mbh-\sigma$ relation.}

\end{itemize}

\noindent For the normal, i.e., non-radio galaxies, one should note that the success of purely
gravitational models does not imply that there are no alternative
models that can fit the data. For example, it cannot be ruled out that
in reality there is a non-gravitational component to the gas motions,
and that the BH masses in these galaxies are below the $M-\sigma$
relation. Some models in the literature, such as those published for
IC 1459 (Cappellari \etal 2002) fall in this category. 
Nevertheless, if one is willing to assume {\it a priori} that the BH
masses of all elliptical galaxies follow the $\Mbh-\sigma$
relation, then one can turn the argument around and conclude that
normal elliptical galaxies do not have a very significant
non-gravitational component in their gas motions. The absence of excess dispersion in non-radio galaxies is corroborated by the fact that the BH mass estimates for this sample by itself yield a relation in good agreement with the $\Mbh-\sigma$ relation (e.g., Figure~\ref{f:mbhdisk}). It is interesting to
note in this context that a similarly good correspondence between BH
masses from gas dispersions and from the $\Mbh-\sigma$
relation was found for non-active early-type {\it spiral} galaxies by
Sarzi et al.~(2002). 

For the low-luminosity radio galaxies in our sample there is also
other evidence, besides the dynamical evidence presented here, for
non-flat gas distributions with turbulent motion. The core
emission-line luminosity of these galaxies correlates with the radio
and optical core luminosity which are both most likely due to
synchrotron emission from the jet (e.g., Chiaberge \etal 1999; Verdoes
Kleijn \etal 2002). This correlation could mean that the emission-line
luminosity is driven by jet photo-ionization. It then implies covering
factors of $\sim 0.3$ and hence thick disks (Capetti
2005). Alternatively, the correlation could indicate that the gas is
excited by shocks induced by jet-gas interactions.

Although it is mostly the radio galaxies that show significant excess
dispersion over a purely gravitational model, there is one normal
galaxy for which this is the case too: IC 989. It should be noted
though that for this particular galaxy the error of $33\kms$ on the
stellar dispersion $\sigma_{\rm s} = 176\kms$ is larger than typical
for the sample. The predicted gas dispersion can be increased by a
factor $\sim 4$ by simply varying $\sigma_{\rm s}$ within its
1-$\sigma$ confidence range. Also, the predicted gas dispersion could
go up by up to a factor $\sim 2$ if the unknown inclination of the
purported gas disk is closer to edge-on than the default value of
$60\deg$ that we assumed. So the evidence for non-gravitational gas
motions in the normal galaxy IC 989 is not strong.

Finally, we note that it is the sample of relatively close to face-on
($i<40\deg$) gas disks which reveal most clearly an excess
dispersion compared to a purely gravitational model. Unfortunately,
this face-on group only contains radio galaxies. However, a statistically significant difference between the excess dispersion in radio and non-radio galaxies remains when one excludes galaxies with $i<40\deg$ gas disks from the analysis. Nevertheless, a good test of
the dichotomy found in this paper will be the modeling of
gas disks at lower inclinations also in non-radio galaxies. Such data is not available currently. 

The evidence for non-gravitational gas motions in radio galaxies makes
the gas dispersion unsuited for dynamical mass estimates in this class
of galaxies. It would be interesting to establish more generally
whether dynamical mass estimates from gas velocities, including
rotation curves, are robust. This would require comparisons of BH
masses from independent indicators (e.g., stellar and gas kinematics)
within the same galaxy. Such a comparison has been performed for a
radio galaxy (NGC 4335, Verdoes Kleijn \etal 2002) and a non-radio
galaxy (Cappellari \etal~2002). Unfortunately, in both cases the
stellar dynamical evidence is too inconclusive to reach firm
conclusions. But it does appear in both cases that the BH mass implied
by thin disk models for the gas rotation curves is lower than would
otherwise have been expected, either from stellar kinematics or from
the $\Mbh-\sigma$ relation. This suggests that the gas might
be moving slower than the circular velocity, as would be expected for
example if there were asymmetric drift in the gas. This occurs for gas
distributions that are dynamically hotter than a thin disk. More
independent BH mass evaluations from gas and other tracers, both for
normal and active galaxies, would be quite valuable to shed more light
on these issues.


\acknowledgments

Support for this work was provided by a grant associated with program GO-9163 from the Space Telescope Science Institute, which is operated by AURA, Inc., under NASA contract NAS5-26555. This paper made use of the LEDA database (see http://leda.univ-lyon1.fr). We thank the anonymous referees for their constructive comments. We thank the first referee in particular for drawing our attention to the dependence of stellar velocity dispersion on galaxy inclination (see the discussion in Section~\ref{s:cave}).

\ifsubmode\else
\baselineskip=10pt
\fi


\clearpage


\ifsubmode\else
\baselineskip=14pt
\fi


\newcommand{\figcapsigmagassigmastar}{\label{f:sigmagassigmastar} HST
measured nuclear gas velocity dispersion versus large-scale
stellar velocity dispersion for our sample galaxies. Open symbols
indicate FRI-type radio galaxies, while solid symbols indicate
galaxies without large-scale radio jets. Most galaxies have nuclear
gas dispersions larger than their stellar dispersions (the equality of
the two being indicated by the line).}

\newcommand{\figcapmbhlit}{\label{f:mbhlit} Observed 
gas velocity dispersion versus predicted gas velocity dispersion for a
thin disk model. We use the BH masses and inclinations inferred from
detailed models for the gas rotation velocities available in the
literature (see Tables~\ref{t:sample} and~\ref{t:data}). Open symbols
indicate FRI-type radio galaxies, while filled circles indicate
galaxies without large-scale radio jets. Circles indicate results from
the thin disk models that neglect the contribution of the stellar mass
to the gravitational potential. Triangles show the predictions when
this simplification is omitted, for the three galaxies for which we
published detailed gas modeling previously. The difference between
the two model approaches is $<10\%$.}

\newcommand{\figcapmbhdisk}{\label{f:mbhdisk} The observed gas
velocity dispersion $\sigma_{\rm g}$ as a function of predicted
velocity dispersion for an infinitely thin gas disk in circular
rotation around a BH with mass $\Mbh$ according to the $\Mbh - \sigma$ relation. Radio galaxies are shown
as open symbols and galaxies without large-scale radio jets
as filled symbols. Squares indicate galaxies for which
the disk inclination is unknown and assumed to be $60\deg$. Galaxies
with disk inclination $i<40\deg$ are indicated by double circles. The typical measurement errors are indicated in the lower
right corner. The dashed and solid lines indicate $\sigma_{\rm
g}$(model)= $(\sqrt{1/2},1,\sqrt{2}) \times \sigma_{\rm g}$(obs),
respectively, to facilitate comparison to the $\Mbh - \sigma$ relation and its $\sim$0.3 dex scatter in BH mass.}

\newcommand{\figcapsigratiodisk}{\label{f:sigratiodisk} 
Ratio of observed and predicted gas dispersions as a function of
gas disk inclination. The predicted
dispersion is assumed to be caused by the circular rotation in a thin
disk around a BH according to the $\Mbh - \sigma$ relation.  The error bars take into account the 0.3 dex intrinsic scatter in
the $\Mbh-\sigma$ relation, and typical errors due
to aperture differences and dispersion measurement errors. The four arrows indicate where points will move if disk inclinations were underestimated within the measurement errors for close to face-on disks. Radio
galaxies (open symbols) have a systematically larger dispersion ratio
than non-radio galaxies (filled symbols). The squares indicate
galaxies for which the dust does not provide a constraint on the gas
disk inclination (placed at fiducial inclination of $60\deg$). If
the assumed disk inclinations are incorrect, and the observed gas is
actually edge-on in all galaxies, then we would have expected the
ratios to lie along the solid line $\sin^{-1} i$. Since many points
lie above this line, inclination errors cannot be the (sole)
explanation of the observed trend.}

\newcommand{\figcapspheroid}{\label{f:spheroid}
The ratio of the velocity dispersion for a collisionless
spheroid divided by that of a thin disk with identical projected
surface density profile, as a function of apparent axis
ratio. The small dots indicate spheroid models with varying
intrinsic axis ratios, powerlaw density slopes and
velocity dispersion anisotropies. The ratios of
observed velocity dispersions and dispersion predicted by thin
disk models from Figure~\ref{f:sigratiodisk} are overplotted. Symbols are as in Fig.~\ref{f:sigratiodisk}.}

\newcommand{\figcapnongrav}{\label{f:nongrav} {\bf Top:} The ratio 
between the observed dispersion and the maximum dispersion expected
from spheroids or thin disks for radio galaxies with
determined inclinations. The ratio is systematically above 1.  {\bf
Middle:} The ratio of non-gravitational dispersion $\sigma_{\rm iso}$
and gravitational dispersion $\sigma_{\rm grav}$ for a two component
gas model which matches the observed dispersion. The model has a
circular thin gas disk which causes $\sigma_{\rm grav}$ measured in
the plane of the disk and a non-gravitational component with an
isotropic gas dispersion $\sigma_{\rm iso}$.  {\bf Bottom:} Similar to
middle plot, but now the non-gravitational dispersion component is
assumed to be perpendicular to the gas disk.}

\newcommand{\figcapagn}{\label{f:agn}
$R_{\rm o,d}$ versus total radio power at 1.4GHz for radio galaxies
(open symbols) and non-radio galaxies (filled symbols). As shown
before, radio galaxies have systematically larger $R_{\rm
disk}$. However, no systematic trend is seen between the
radiative power of the AGN at radio wavelengths and the excess
dispersion within the radio galaxy sample.}

\newcommand{\figcaphalflight}{\label{f:halflight} {\bf Top:} angular 
radius inside which half of the total gas emission is
contained versus distance for radio galaxies (open circles) and
non-radio galaxies (filled circles). The half light radius of the PSF
for these observations is $\sim 0.05''$. {\bf Bottom:} same as top
panel, but now the half light radius is expressed in parsec.}

\newcommand{\figcapmagsigma}{\label{f:magsigma} Stellar velocity dispersion as a function of host magnitude for the sample galaxies. Galaxies with dust disks with $i<40\deg$ are indicated by a double circle.}



\ifsubmode 
\figcaption{\figcapsigmagassigmastar} \clearpage
\figcaption{\figcapmbhlit} \clearpage
\figcaption{\figcapmbhdisk} \clearpage
\figcaption{\figcapsigratiodisk} \clearpage
\figcaption{\figcapspheroid} \clearpage
\figcaption{\figcapnongrav} \clearpage
\figcaption{\figcapagn} \clearpage
\figcaption{\figcaphalflight} \clearpage
\figcaption{\figcapmagsigma} \clearpage
\else\printfigtrue\fi

\ifprintfig


\clearpage
\begin{figure}
\epsscale{1.0}
\plotone{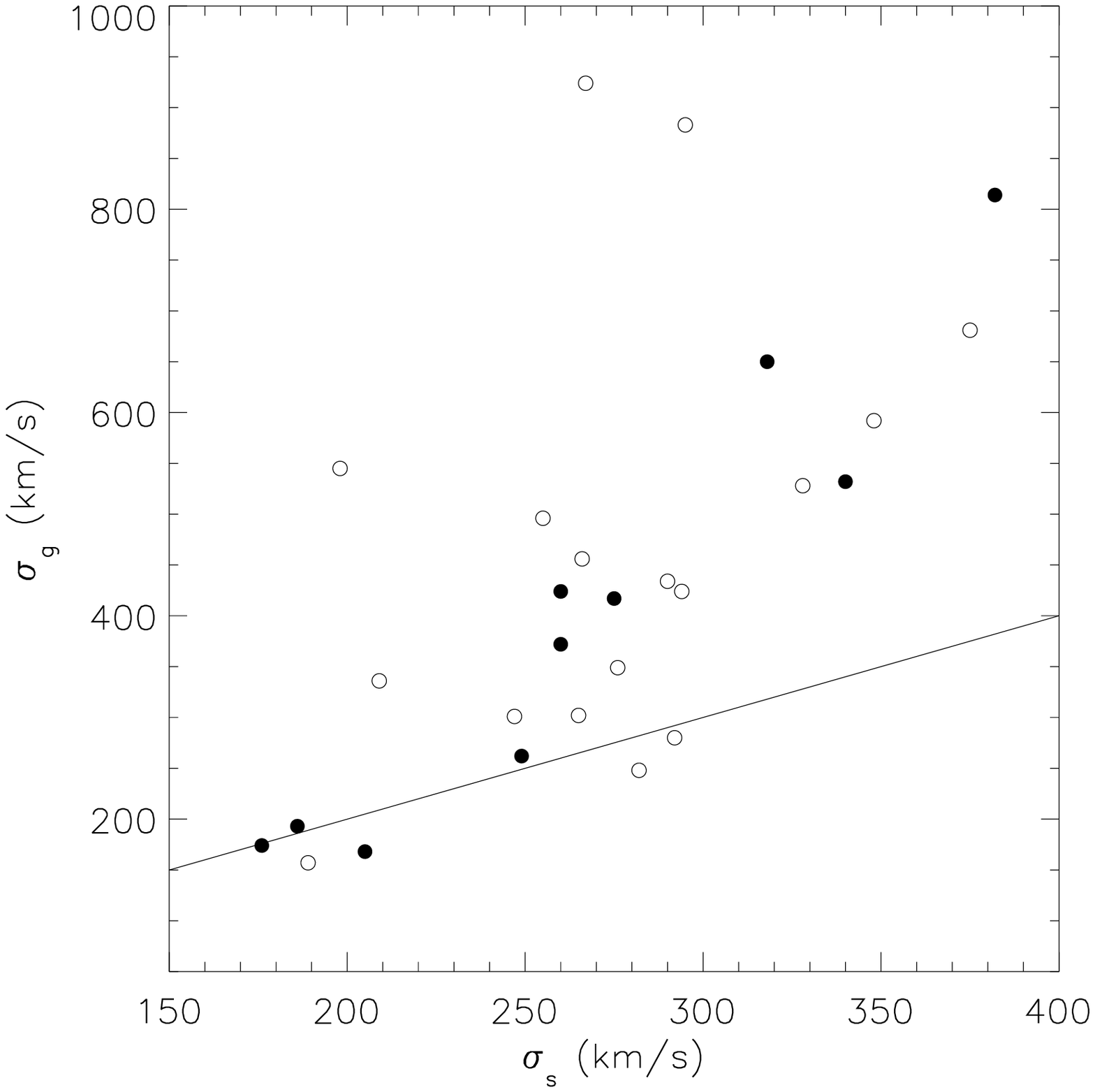}
\ifsubmode
\vskip3.0truecm
\setcounter{figure}{0}
\addtocounter{figure}{1}
\centerline{Figure~\thefigure}
\else\figcaption{\figcapsigmagassigmastar}\fi
\end{figure}
 
\clearpage
\begin{figure}
\epsscale{1.0}
\plotone{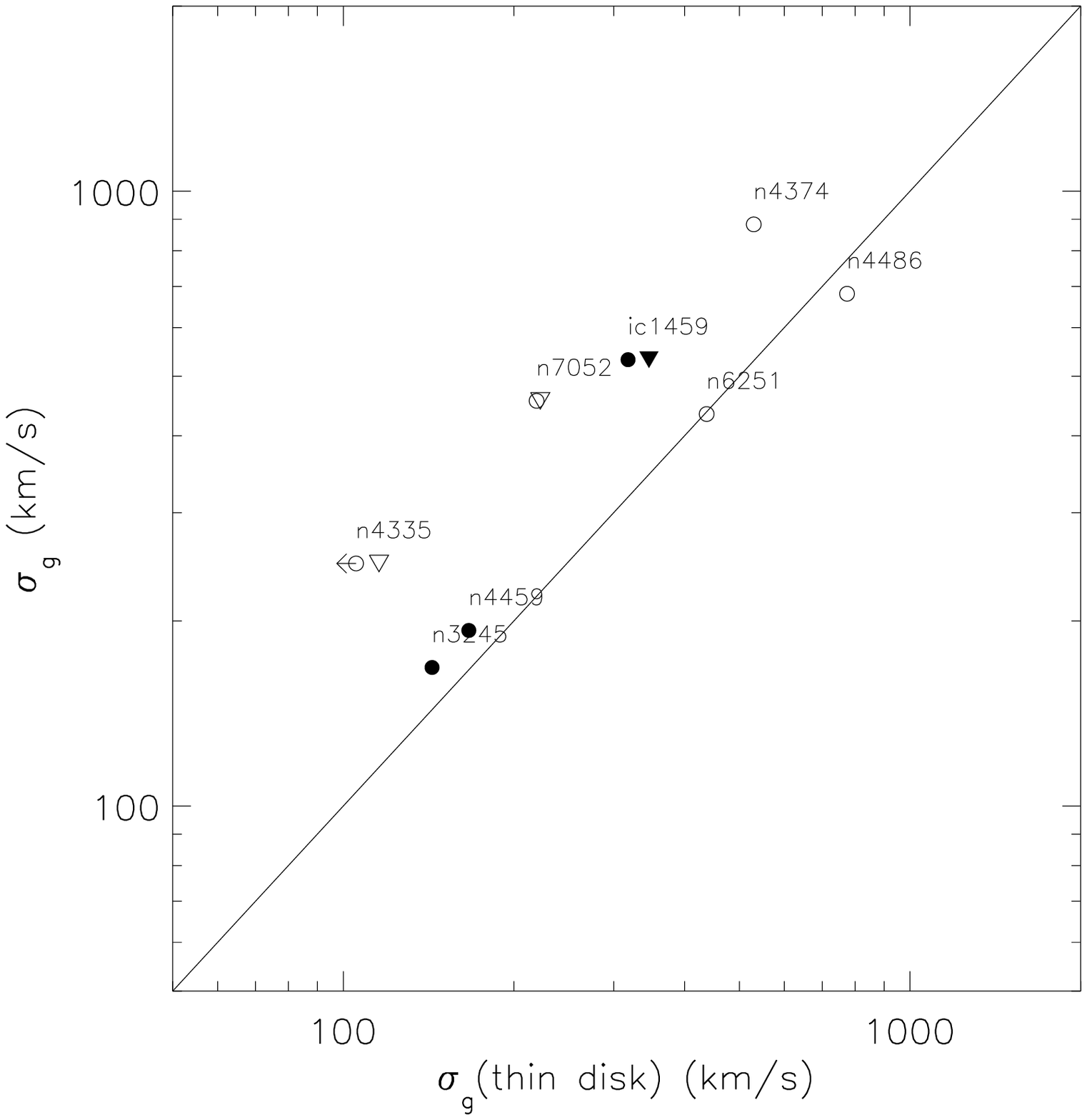}
\ifsubmode
\vskip3.0truecm
\addtocounter{figure}{1}
\centerline{Figure~\thefigure}
\else\figcaption{\figcapmbhlit}\fi
\end{figure}
 
\clearpage
\begin{figure}
\epsscale{1.0}
\plotone{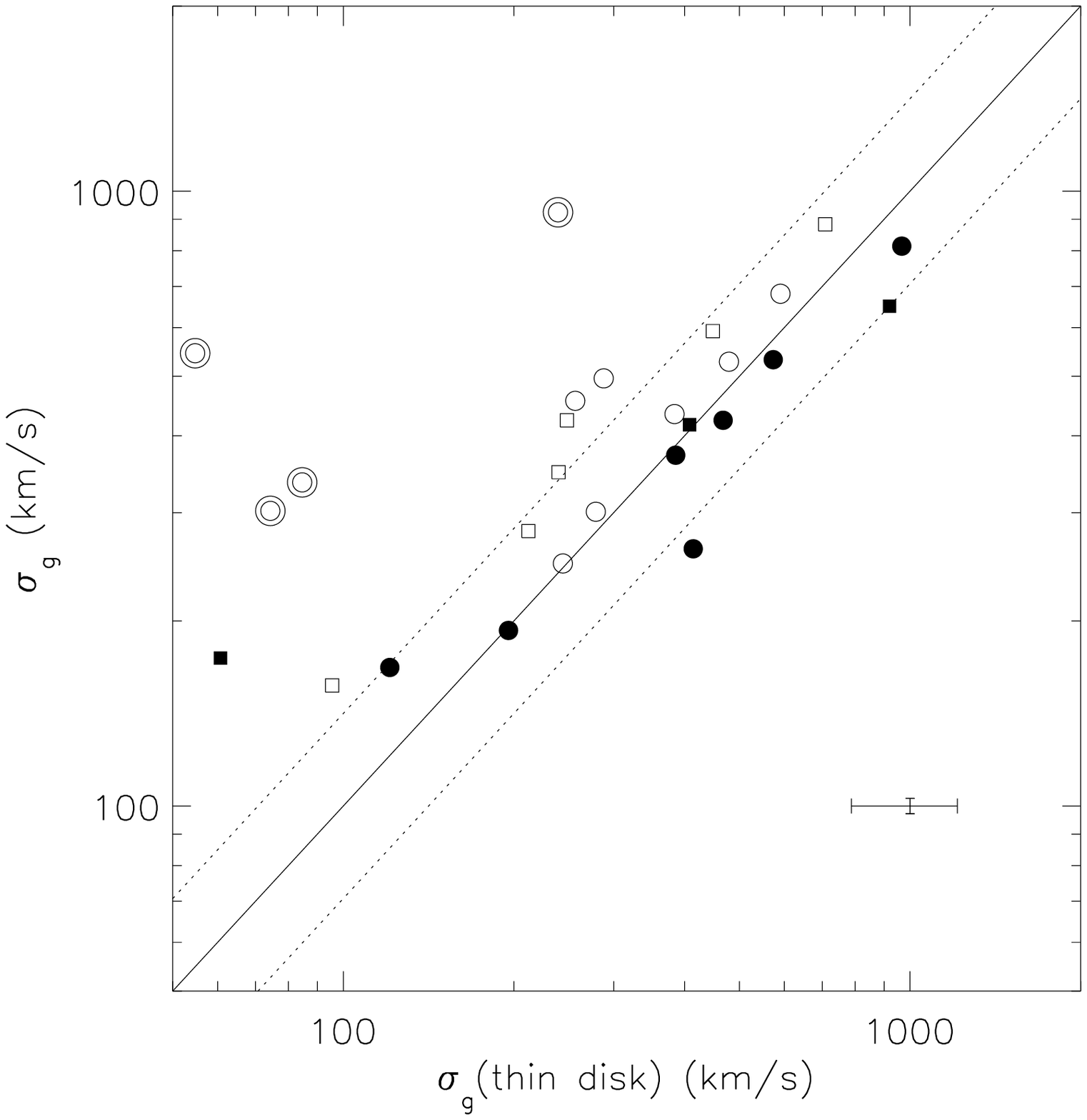}
\ifsubmode
\vskip3.0truecm
\addtocounter{figure}{1}
\centerline{Figure~\thefigure}
\else\figcaption{\figcapmbhdisk}\fi
\end{figure}

\clearpage
\begin{figure}
\epsscale{1.0}
\plotone{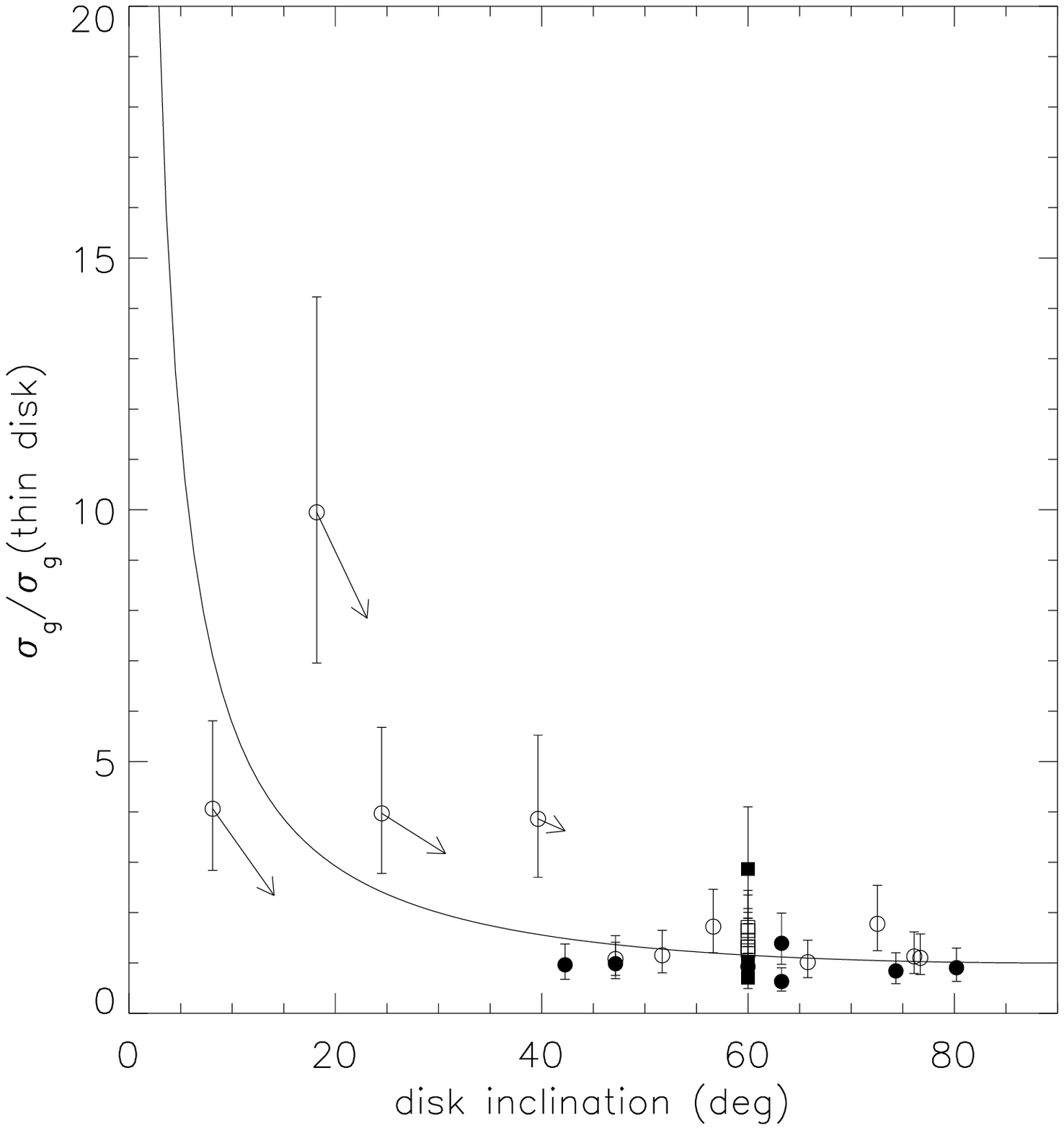}
\ifsubmode
\vskip3.0truecm
\addtocounter{figure}{1}
\centerline{Figure~\thefigure}
\else\figcaption{\figcapsigratiodisk}\fi
\end{figure}

\clearpage
\begin{figure}
\epsscale{1.0}
\plotone{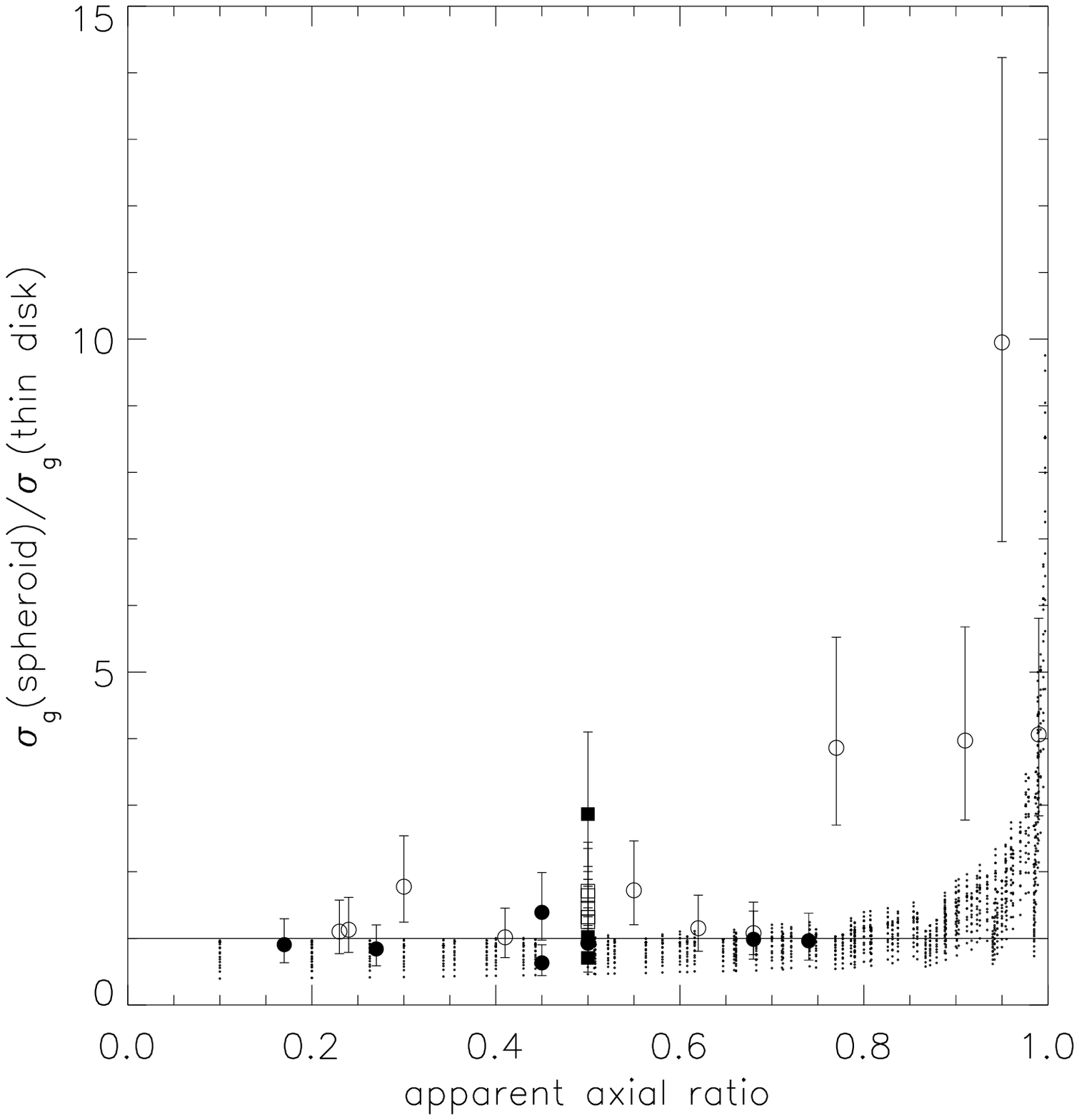}
\ifsubmode
\vskip3.0truecm
\addtocounter{figure}{1}
\centerline{Figure~\thefigure}
\else\figcaption{\figcapspheroid}\fi
\end{figure}

\clearpage
\begin{figure}
\epsscale{1.0}
\plotone{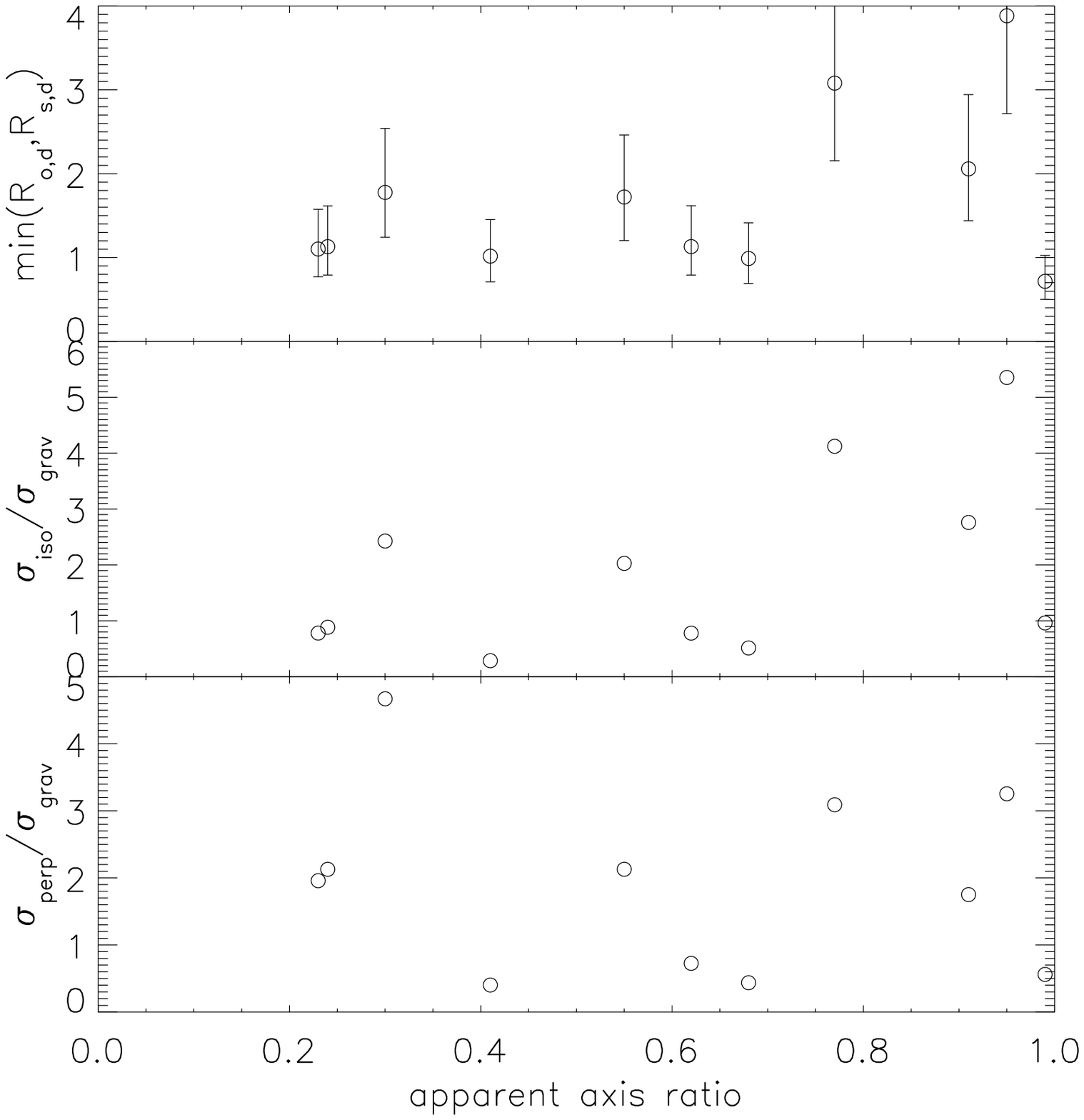}
\ifsubmode
\vskip3.0truecm
\addtocounter{figure}{1}
\centerline{Figure~\thefigure}
\else\figcaption{\figcapnongrav}\fi
\end{figure}

\clearpage
\begin{figure}
\epsscale{1.0}
\plotone{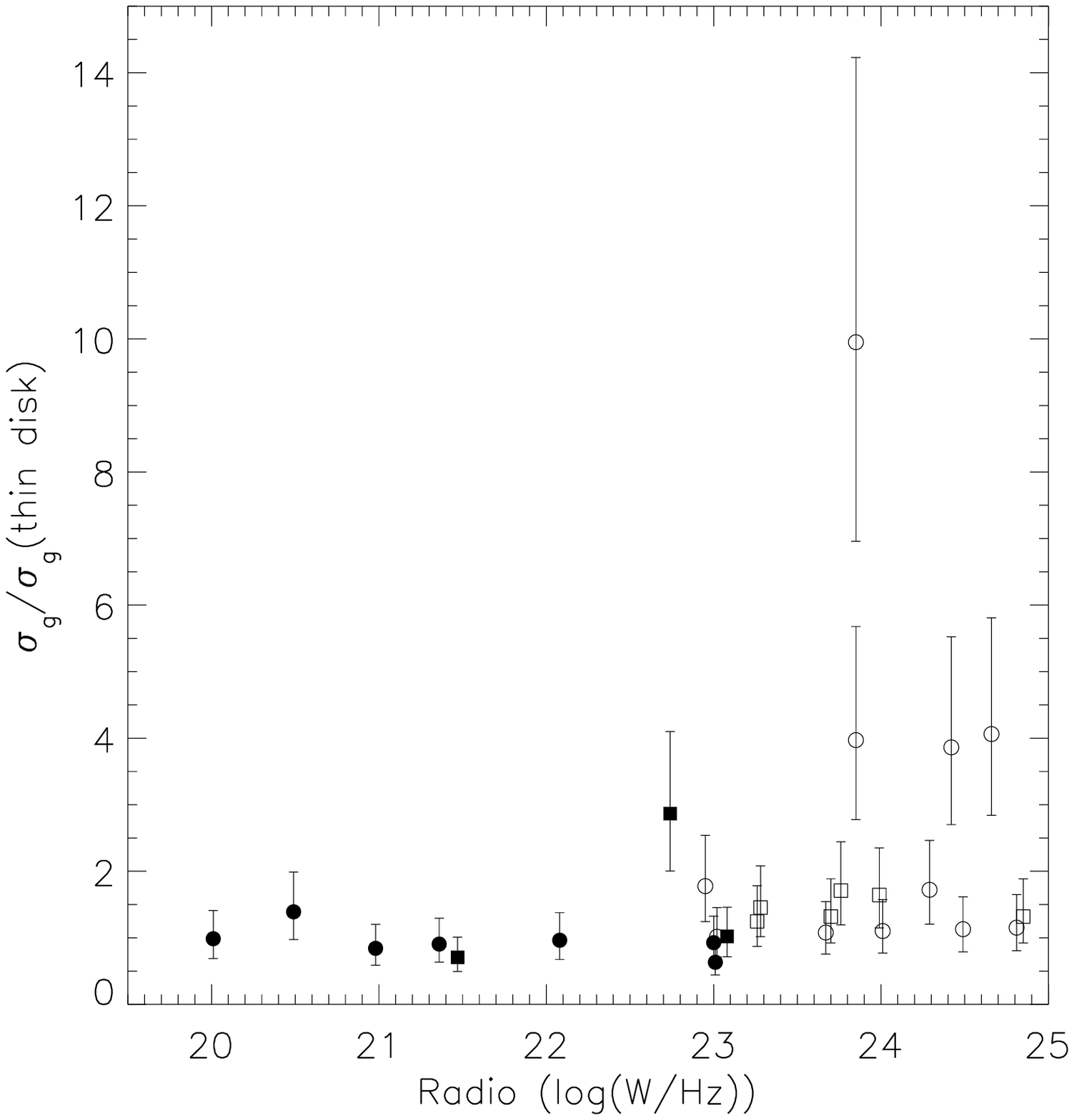}
\ifsubmode
\vskip3.0truecm
\addtocounter{figure}{1}
\centerline{Figure~\thefigure}
\else\figcaption{\figcapagn}\fi
\end{figure}
 
\clearpage
\begin{figure}
\epsscale{1.0}
\plotone{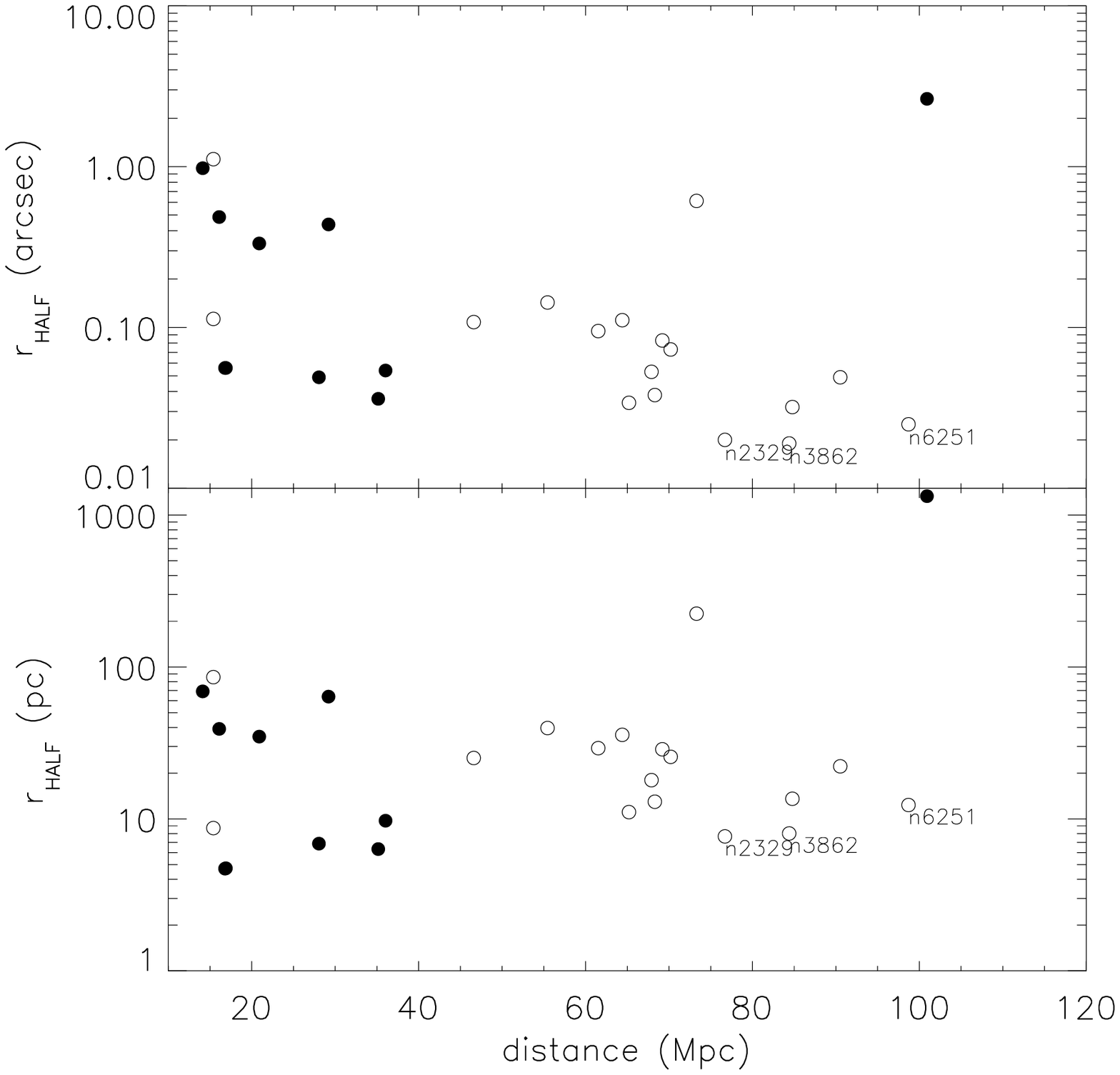}
\ifsubmode
\vskip3.0truecm
\addtocounter{figure}{1}
\centerline{Figure~\thefigure}
\else\figcaption{\figcaphalflight}\fi
\end{figure}
 
\clearpage
\begin{figure}
\epsscale{1.0}
\plotone{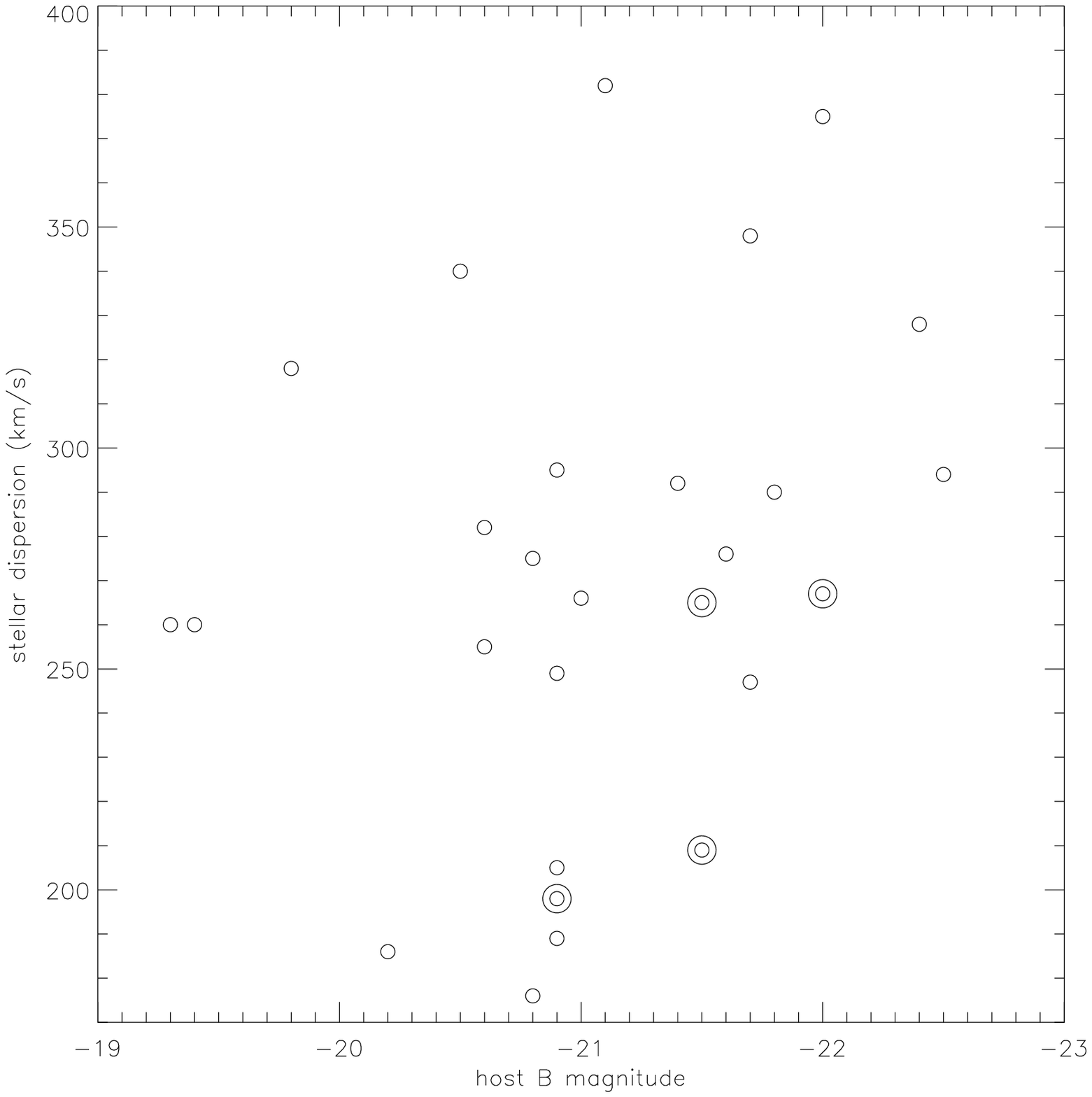}
\ifsubmode
\vskip3.0truecm
\addtocounter{figure}{1}
\centerline{Figure~\thefigure}
\else\figcaption{\figcapmagsigma}\fi
\end{figure}
 

\fi


\clearpage
\ifsubmode\pagestyle{empty}\fi


\begin{deluxetable}{llrrrrrrrl}
\tabletypesize{\small}
\setcounter{table}{0}
\tablewidth{0pt}
\tablecaption{Early-Type Galaxy Sample\label{t:sample}}
\tablehead{ \colhead{Galaxy} & \colhead{Type} & \colhead{FRI} & \colhead{D} & \colhead{$M_B$} & \colhead{$\sigma_{\rm s}$} & \colhead{$\sigma_{\rm g}$} & \colhead{$\Mbh$} & \colhead{$L_{\rm radio}$} & \colhead{ref} \\ 
\colhead{} & \colhead{} & \colhead{} & \colhead{(Mpc)} & \colhead{(mag)} & \colhead{(km/s)} & \colhead{(km/s)} & \colhead{($10^8\Msun$)} & \colhead{log(W/Hz)} & \colhead{} \\ 
\colhead{(1)} & \colhead{(2)} & \colhead{(3)} & \colhead{(4)} & \colhead{(5)} & \colhead{(6)} & \colhead{(7)} & \colhead{(8)} & \colhead{(9)} & \colhead{(10)} \\ } 
\startdata 
IC 989    & E    &   & 101 & -20.8 & 176     & 174$^*$:    & ...  & 22.74 & r3\\ 
IC 1459   & E    &   &  29 & -20.5 & 340$^b$ & 532         &  1.3 & 23.00 & b1,r4\\ 
NGC 315   & E    & 1 &  68 & -22.4 & 328     & 528$^2$     & ...  & 24.01 & r1\\  
NGC 383   & E/SO & 1 &  65 & -22.0 & 267     & 924$^2$     & ...  & 24.42 & r1\\ 
NGC 541   & E/SO & 1 &  73 & -21.5 & 209     & 336         & ...  & 23.85 & r1\\ 
NGC 741   & E    & 1 &  70 & -22.5 & 294     & 424         & ...  & 23.76 & r1\\ 
NGC 2329  & E/SO & 1 &  77 & -21.7 & 247     & 301         & ...  & 23.67 & r1\\ 
NGC 3078  & E    &   &  35 & -20.9 & 249     & 262$^*$     & ...  & 23.01 & r5\\ 
NGC 3245  & SO   &   &  21 & -20.0 & 205$^b$ & 168$^1$     &  2.1 & 20.49 & b2, r6\\ 
NGC 3862  & E    & 1 &  84 & -21.5 & 265     & 302         & ...  & 24.66 & r1\\ 
NGC 3998  & SO   &   &  14 & -19.8 & 318     & 650$^{*1}$  & ...  & 21.47 & r7\\ 
NGC 4278  & E    &   &  16 & -19.3 & 260     & 372$^{*1}$  & ...  & 22.08 & r6 \\ 
NGC 4335  & E    & 1 &  62 & -20.6 & 282$^c$ & 248$^2$     &$<1.0$& 23.02 & b3,r1\\ 
NGC 4374  & E    & 1 &  15 & -20.9 & 295     & 883         &  3.6 & 23.26 & b4,r1\\
NGC 4459  & SO   &   &  17 & -20.2 & 186$^b$ & 193         &  0.7 & 20.01 & r3\\ 
NGC 4486  & E    & 1 &  15 & -22.0 & 375$^b$ & 681:$^2$    & 29.0 & 24.81 & b6,r1\\
NGC 4526  & SO   &   &  36 & -19.4 & 260     & 424$^*$     & ...  & 21.36 & r3\\ 
NGC 5077  & E    &   &  36 & -20.8 & 275     & 417$^*$     & ...  & 23.08 & r5 \\ 
NGC 5127  & E    & 1 &  64 & -20.9 & 189     & 157         & ...  & 23.99 & r1\\ 
NGC 5490  & E    & 1 &  69 & -21.4 & 292     & 280$^2$     & ...  & 23.70 & r1\\ 
NGC 6251  & E    & 1 &  99 & -21.8 & 290$^b$ & 434$^1$:    &  7.8 & 24.49 & b7,r7\\ 
NGC 6861  & E/SO &   &  28 & -21.1 & 382$^d$ & 814$^*$     & ...  & 20.98 & r8\\  
NGC 7052  & E    & 1 &  55 & -21.0 & 266$^b$ & 456         &  3.1 & 22.95 & b8,r1\\ 
NGC 7626  & E    & 1 &  47 & -21.6 & 276     & 349$^2$     & ...  & 23.28 & r1\\ 
UGC 1841  & E    & 1 &  85 & -21.7 & 348     & 592$^2$     & ...  & 24.85 & r1\\ 
UGC 7115  & E    & 1 &  91 & -20.9 & 198     & 545:        & ...  & 23.85 & r1\\ 
UGC 12064 & E/SO & 1 &  68 & -20.6 & 255     & 496$^2$     & ...  & 24.29 & r1\\ 
\enddata 
\tablecomments{\scriptsize General properties of the galaxy sample. Col.(2): Hubble 
classification from the LEDA catalogue
(http://leda.univ-lyon1.fr/). Col.(3): galaxies indicated by a 1
contain large scale Fanaroff \& Riley type 1 (1974)
radio-jets. Col.(4): distances from Faber et al. (1989), Tonry et
al. (2001), or, if not available, directly from observed recession
velocity and $H_0$ = 75kms$^{-1}$Mpc$^{-1}$. Col.(5): Absolute blue
magnitude from LEDA. Col.(6): central stellar velocity dispersions
typically integrated over an aperture area of several arcsec$^2$. The
dispersions are from the LEDA catalogue except for: $^b$: Tremaine
\etal ( 2002); $^c$: Verdoes Kleijn \etal (2002); $^d$: from Koprolin
\& Zeilinger (2000). See Section~\ref{s:sampledata} for discussion of the errors. Col.(7): gas velocity dispersions of narrow
{\HalphaNII} and [SII] emission-lines as measured from HST spectra at
the central aperture. See Section~\ref{s:sampledata} for discussion of the errors. These measurements are taken from the literature
or obtained by us (indicated by $^*$). A $^1$ indicates that a broad
H$\alpha$ line was fitted in addition to the narrow-lines of the
H$\alpha$+[NII] complex. A $^2$ indicates that a broad H$\alpha$ line
is perhaps present (cf.~Section~\ref{s:cave}). A ":" indicates
galaxies for which a larger dispersion is measured just outside the
central aperture (see Section~\ref{s:sampledata} for
details). Col.(8): black hole masses from detailed dynamical gas disk
modeling reported in the literature. Col.(9): total radio luminosities
at 1.4 GHz (assuming a spectral index $\alpha = 0.75$ for $f_\nu \sim
\nu^{-\alpha}$). Col.(11): BH mass references: b1: Cappellari \etal
(2002); b2: Barth \etal (2001); b3: Verdoes Kleijn \etal (2002); b4:
Bower \etal (1998); b5: Sarzi \etal (2001); b6: Macchetto \etal
(1997); b7: Ferrarese \& Ford (1999); b8: van der Marel \& van den
Bosch (1998). Radio emission references: r1: Condon \& Broderick
(1988); r3: Dressel \& Condon (1978); r4: Wright \& Otrupcek (1990);
r5: Griffith \etal (1994); r6: Becker, White \& Helfand (1995); r7:
White \& Becker (1992); r8: Mauch \etal~(2003).}
\end{deluxetable}


\begin{deluxetable}{lrrlllll}
\tabletypesize{\small}
\tablewidth{0pt}
\tablecaption{Spectroscopic properties and results\label{t:data}}
\tablehead{
\colhead{Galaxy} & \colhead{program} & \colhead{ap} & \colhead{$R_1$} & \colhead{$R_2$} & \colhead{$I_1/I_2$} & \colhead{$i$} & \colhead{ref} \\ 
\colhead{} & \colhead{} & \colhead{($('')^2$)} & \colhead{($''$)} & \colhead{($''$)} & \colhead{} & \colhead{($\deg$)} &  \colhead{} \\
\colhead{(1)} & \colhead{(2)} & \colhead{(3)} & \colhead{(4)} & \colhead{(5)} & \colhead{(6)} & \colhead{(7)} & \colhead{(8)} \\
}
\startdata
IC 989     & 7354 & 0.1$\times$0.05    & 3.3$\times 10^{-2}$ & 1.6$\times 10^{ 0}$ & 2.2$\times 10^{-1}$ & 60$^a$ & \\ 
IC 1459    & 7352 & 0.1$\times$0.05    & 9.0$\times 10^{-3}$ & 2.7$\times 10^{-1}$ & 3.6$\times 10^{-2}$ & 60     & 1 \\ 
NGC 315    & 8236 & 0.1$\times$0.05    & 1.7$\times 10^{-2}$ & 2.0$\times 10^{-1}$ & 4.7$\times 10^{-3}$ & 77     & 5\\  
NGC 383    & 8236 & 0.1$\times$0.05    & 2.0$\times 10^{-2}$ & 2.1$\times 10^{-2}$ & 3.8$\times 10^{-2}$ & 40     & 5\\ 
NGC 541    & 8236 & 0.2$\times$0.1     & 1.4$\times 10^{-2}$ & 5.3$\times 10^{-1}$ & 1.9$\times 10^{-3}$ & 25     & 5\\ 
NGC 741    & 8236 & 0.2$\times$0.1     & 4.3$\times 10^{-2}$ & 4.4$\times 10^{-3}$ & 4.4$\times 10^{-2}$ & 60$^a$ & 5\\ 
NGC 2329   & 8236 & 0.2$\times$0.1     & 1.2$\times 10^{-2}$ & 8.9$\times 10^{-3}$ & 3.2$\times 10^{-2}$ & 47     & 5\\
NGC 3078   & 9163 & 0.2$\times$0.1     & 2.0$\times 10^{-2}$ & 3.2$\times 10^{-1}$ & 4.0$\times 10^{-4}$ & 63     & \\ 
NGC 3245   & 7403 & 0.2$\times$0.05    & 2.0$\times 10^{-1}$ & 6.4$\times 10^{-2}$ & 1.6$\times 10^{-1}$ & 63     & 2 \\ 
NGC 3862   & 8236 & 0.2$\times$0.1     & 1.1$\times 10^{-2}$ & 2.0$\times 10^{-4}$ & 3.7$\times 10^{-2}$ & 08     & 5\\
NGC 3998   & 7354 & 0.1$\times$0.05    & 1.6$\times 10^{-2}$ & 6.2$\times 10^{-1}$ & 8.2$\times 10^{-3}$ & 60$^a$ &  \\
NGC 4278   & 7403 & 0.2$\times$0.05    & 2.6$\times 10^{-2}$ & 3.8$\times 10^{-1}$ & 1.7$\times 10^{-2}$ & 42     &  \\
NGC 4335   & 8236 & 0.2$\times$0.1     & 5.7$\times 10^{-2}$ & 5.8$\times 10^{-2}$ & 4.4$\times 10^{-2}$ & 66     & 5\\ 
NGC 4374   & 7124 & 0.2$\times$0.05    & 2.1$\times 10^{-2}$ & 7.3$\times 10^{-1}$ & 8.7$\times 10^{-3}$ & 60$^a$ & 5\\ 
NGC 4459   & 7361 & 0.2$\times$0.25    & 2.2$\times 10^{-2}$ & 9.2$\times 10^{-2}$ & 3.4$\times 10^{-2}$ & 47     & 3 \\ 
NGC 4486   & 8666 & 0.2$\times$0.05    & 6.7$\times 10^{-2}$ & 2.2$\times 10^{-3}$ & 4.6$\times 10^{-2}$ & 52     & 5\\ 
NGC 4526   & 9163 & 0.2$\times$0.05    & 3.3$\times 10^{-2}$ & 1.1$\times 10^{-2}$ & 3.6$\times 10^{-2}$ & 80     & \\ 
NGC 5077   & 7354 & 0.1$\times$0.05    & 3.2$\times 10^{-2}$ & 3.1$\times 10^{-2}$ & 3.9$\times 10^{-2}$ & 60$^a$ & \\ 
NGC 5127   & 8236 & 0.2$\times$0.1     & 6.4$\times 10^{-2}$ & 9.4$\times 10^{-2}$ & 5.3$\times 10^{-2}$ & 60$^a$ & 5\\ 
NGC 5490   & 8236 & 0.2$\times$0.1     & 5.0$\times 10^{-2}$ & 1.4$\times 10^{-3}$ & 4.2$\times 10^{-2}$ & 60$^a$ & 5\\ 
NGC 6251   & 6653 & 0.1$\times$0.1     & 1.0$\times 10^{-2}$ & 4.1$\times 10^{-2}$ & 3.1$\times 10^{-2}$ & 32     & 4 \\ 
NGC 6861   & 9163 & 0.2$\times$0.1     & 2.9$\times 10^{-2}$ & 1.0$\times 10^{-4}$ & 3.8$\times 10^{-2}$ & 74     & \\ 
NGC 7052   & 8236 & 0.1$\times$0.05    & 2.3$\times 10^{-2}$ & 1.7$\times 10^{-1}$ & 2.9$\times 10^{-2}$ & 73     & 5\\ 
NGC 7626   & 8236 & 0.2$\times$0.1     & 2.3$\times 10^{-2}$ & 7.3$\times 10^{-2}$ & 7.0$\times 10^{-1}$ & 60$^a$ & 5\\ 
UGC 1841   & 8236 & 0.2$\times$0.1     & 1.9$\times 10^{-2}$ & 1.5$\times 10^{-2}$ & 4.0$\times 10^{-2}$ & 60$^a$ & 5\\ 
UGC 7115   & 8236 & 0.2$\times$0.1     & 1.8$\times 10^{-2}$ & 7.0$\times 10^{-2}$ & 4.8$\times 10^{-2}$ & 18     & 5\\ 
UGC 12064  & 8236 & 0.1$\times$0.05    & 2.3$\times 10^{-2}$ & 2.6$\times 10^{-3}$ & 3.9$\times 10^{-2}$ & 57     & 5\\ 
\enddata
\tablecomments{\small Col.(2): HST program number of spectral observations. 
Col.(3): size of central aperture through which $\sigma_{\rm g}$ was
measured. The spectra are obtained using the HST/STIS spectrograph
with the G750M grating except for IC 1459 for which the G430L grating
was used and NGC 6251 for which the HST/FOS spectrograph with the
G780H grating was used. Col.(4)-(6): parameters $R_1$, $R_2$ and
$I_2/I_1$ of the unconvolved emission-line surface brightness profile
$I(R)=I_1 \exp(-R/R_1) + I_2 \exp(-R/R_2)$ fitted to the
observations. Col.(7): inclination used in disk modeling. The
inclination has been determined by us or other authors from larger
scale dust disks present in sources except those indicated with a
$^a$. For these sources an $i=60\deg$ was assumed (see
Section~\ref{s:sampledata} for details). Col.(9): references for the
galaxies for which we used flux measurements (and gas velocity
dispersions) reported in the literature. 1: Cappellari \etal (2002);
2: Barth \etal (2001); 3: Sarzi \etal (2001); 4: Ferrarese \& Ford
(1999); 5: Noel-Storr \etal (2003).  }
\end{deluxetable}



\end{document}